\documentstyle[aps,graphicx,floats,tighten,preprint]{revtex}

\def\be{\begin{equation}}
\def\ee{\end{equation}}
\def\bea{\begin{eqnarray}}
\def\eea{\end{eqnarray}}
\def\Bphi{\mbox{\boldmath $\Phi$}}

\def\hphi{\mbox{\boldmath $\hat\Phi$}}
\def\Bx{\mbox{\boldmath $x$}}
\def\Bxi{\mbox{\boldmath $\xi$}}
\def\Bk{\mbox{\boldmath $k$}}
\begin{document}
\title{DCC Dynamics in (2+1)D-O(3) model}
\author{G. Holzwarth\thanks{%
e-mail: holzwarth@physik.uni-siegen.de}}
\address{Fachbereich Physik, Universit\"{a}t Siegen, 
D-57068 Siegen, Germany} 
\maketitle

\begin{abstract}\noindent
The dynamics of symmetry-breaking after a quench is numerically
simulated on a lattice for the (2+1)-dimensional $O(3)$ model. In 
addition to the standard sigma-model with temperature-dependent
$\Phi^4$-potential the energy 
functional includes  a four-derivative current-current coupling to
stabilize the size of the emerging extended topological textures. 
The total winding number can be conserved by constraint. 
As a model for the chiral phase
transition during the cooling phase after a hadronic collision 
this allows to investigate the interference of 'baryon-antibaryon'
production with the developing disoriented aligned domains. 
The growth of angular correlations, condensate, average orientation
is studied in dependence of texture size, quench rate, symmetry
breaking. The classical dissipative dynamics determines
the rate of energy emitted from the relaxing source for each component
of the 3-vector field which provides a possible signature for
domains of Disoriented Chiral Condensate. We find that the 'pions'
are emitted in two distinct pulses; for sufficiently small lattice size
the second one carries the DCC signal, but it is strongly suppressed as
compared to simultaneous 'sigma'-meson emission. We compare the
resulting anomalies in the distributions of DCC pions with probabilities
derived within the commonly used coherent state formalism.
\end{abstract} 

\vspace{3cm}
\leftline{PACS numbers: 11.27.+d;12.39.Dc,Fe;25.75;64.60.Cn;75.10-b;75.40.Mg}  

\leftline{Keywords: Chiral Phase Transition, Disoriented
Chiral Condensate, Sigma models,} 
\leftline{Topological textures, Skyrmions, Bags}

\newpage

\section{Introduction}

The observation of clear signatures for the chiral phase transition
still remains one of the outstanding challenges of hadron physics.
There is hope that the extreme energy densities located inside the spatial
region between receding baryonic slabs shortly after ultrarelativistic
baryon or heavy-ion collision may provide physical conditions for such
a phase transition to take place.

From the theoretical point of view we are dealing with a
complex dynamical system, strongly interacting quantum
fields with many different fermionic and bosonic degrees of freedom,
far from equilibrium, creating and emitting baryons and mesons
from a rapidly expanding and cooling spatial volume. 
Correspondingly, there is a vast theoretical literature ranging
from cascade models in the colored partonic degrees of freedom
to hydrodynamic flow models for hadronic currents, which might be
applicable at different stages and shed light on different aspects
during the time evolution of such events. 
If within the hot fireball chiral symmetry indeed is restored 
then the dynamics of its spontaneous breaking and formation of a chiral
condensate is one important aspect of the cooling process in which
the highly excited spatial volume returns back to normal physical
vacuum. 

A possible strategy which has proven useful for many other 
cases of phase transitions in microscopically very complex systems
is to comprise the essential phenomenology into an effective action
for an ordering parameter field. For the hadron phenomenology 
near $T=0$ a chiral field with spontaneously broken
$O(4)$-symmetry has become a standard and successful
tool, so it is natural to try to extend and apply this concept to
the temperature region where the chiral phase transition is expected.
In its most simple version the effective action then consists of the
(3+1)dimensional $O(4)$ linear $\sigma$-model, with a suitable
temperature-dependent $\Phi^4$-potential for the spontaneous symmetry
breaking. Depending on the ratio between relaxation and cooling
times, the dynamics of the ordering field then follows the
cooling quench imposed on the temperature dependence of the potential.
As in other cases of multi-component field ordering, it is characterized by 
formation and growth of misaligned domains separated by domain
walls or other topologically nontrivial structures, depending on field
and space dimensionality~\cite{Bray}. Numerical simulations in this
classical framework have 
been performed and they confirm the transient formation of domains of
false vacuum, in which the condensate is approaching its vacuum value
while the direction of the aligned field still deviates from the
surrounding vacuum~\cite{Raja}. 
The possible influence of quantum- and thermal effects has been
investigated, mostly in mean-field approximation~\cite{Randrup}.
It has been suggested that pions
emitted from these domains of disoriented chiral condensate (DCC) carry
the misalignment in their isospin multiplicities, and in this way may
provide a signature of the phase transition~\cite{Anselm}.

A peculiar feature of the chiral $O(4)$-field in 3 spatial dimensions
is that its winding density may be identified with baryon
density~\cite{Skyrme}. This 
has led to a most interesting and remarkably successful model for
baryon structure and dynamics at $T=0$. The corresponding 
topological arguments~\cite{Kibble} have been used~\cite{Ellis} to
estimate nascent baryon-antibaryon 
multiplicities in random chiral field configurations at high temperature;
of course, the first few time steps in an ordering evolution lead
to an almost instantaneous reduction of the initially large and rapidly
fluctuating winding densities.

Strictly, the topological conservation of the total winding number is
limited to a field manifold where the point $\Phi=0$ is excluded. 
In numerical simulations of field evolutions on a discrete
spatial lattice, however, even 
topological conservation laws have to be enforced by constraint, because 
arguments based on continuity cannot be applied. Therefore there is no
principal difference for implementing on a lattice the conservation of winding
number in the nonlinear or linear realization of the $O(4)$-symmetry.
This allows to study formation of condensate and disoriented domains
during the phase transition with simultaneously forming baryons plus
antibaryons such that the total baryon number is fixed to any desired
value. It is well known that topological textures play an important
role in ordering transitions, so it is of special interest to study how
the stabilization of extended baryonic structures and baryon number
conservation may affect the growth laws of DCC domains, and possible DCC
signatures in the emitted radiation. 
For this purpose it is necessary to include in the effective action 
terms which stabilize the spatial size of the baryons
generated in the course of the ordering evolution. Scaling arguments
show that an additional four-derivative term is
sufficient for soliton stabilization in 2- and 3-dimensional $O(n)$
models. A different extension, the inclusion of the anomaly term, 
has been advocated to increase the probability of DCC-formation~\cite{Asa}.

The linear $\sigma$-model (even with explicite inclusion of
quarks) sometimes has been used~\cite{Biro} to investigate 
the possibility of DCC-signals by considering spatially averaged
fields, only. Naturally, within such an approximation the
effects we are investigating here, are lost. On the other hand, there
is no immediate need to explicitly include fermionic degrees of freedom
in our present approach.

Before embarking on a full-scale simulation of the (3+1)-dimensional
$O(4)$ case it is helpful to analyse the relevant questions in the
(2+1)-dimensional $O(3)$ model~\cite{Zak}. It shares all essentials with the 
higher-dimensional case. Although it represents an efficient tool for
the description of 2-dimensional spin systems in its own
right~\cite{Sondhi}, we 
here discuss it mainly in view of its hadron physics generalization.
That is, we denote the order parameter field as 'chiral' meson field, we
call the fluctuations 'pions' with internal 'isospin' components,
and $\sigma$-mesons, which aquire mass through spontaneous breaking
of the chiral $O(3)$-symmetry, and we call the domains with small
values of the condensate and unbroken symmetry 'bags', in which the
'baryonic' winding density is confined. Although there is no 
compelling connection to fix the parameters of the model from its
higher-dimensional analogue, we choose them in correspondence to
simulate relative sizes and masses for baryons and mesons.
Various features of the static solutions of this
model, and their formation in relaxation processes after a quench
have been studied in~\cite{Ho,HoKl}. In the present investigation we discuss
in more detail the different stages which characterize the long-time
ordering evolution under dissipative dynamics, the dependence of the
relevant times on model parameters, different types of symmetry
breaking, and quench rates. Our main aim, however, is an attempt to
extract from the dynamics of the evolution the possibility of a
DCC-signal. Generally, the emission of radiation from a moving source
is determined by quantum-field-theoretic amplitudes which require
a separation of the time-dependent field into a classically moving
part and quantum fluctuations. The result is commonly put into the
statement that the 'intensity of pion radiation is proportional
to the square of the classical pion field'~\cite{Anselm}. 
This argument is the basis
of the interest in DCC-pions. We analyse the validity of this statement
under the assumption of dissipative dynamics, i.e. assuming that the 
energy loss of the relaxing and ordering classical configurations
is carried away by propagating fluctuations. Under this assumption
we obtain the timing and strength and isospin distribution of DCC-pions
from purely classical considerations.

This general dynamical scheme is outlined in sect.2, while the
specifics of the model and the observables of interest are defined in
sect.3. The essential features of configurations during the course of
the transition, the growth of angular correlations, decrease of numbers
of textures, saturation of condensate and average orientation,
formation of baryonic clusters, are presented in chpts.4,5, and
6, in dependence on parameters of the model, symmetry breaking, and
cooling rates. Finally, in sect.7, we analyse the rates of energy loss
with respect to possible DCC pion pulses, and extract the isospin
distributions from large ensembles of events. Conclusions are drawn
in sect.8.

\section{Nonequilibrium field dynamics and DCC signals}

We consider an effective action for an $O(3)$ vector field $\Bphi$
\be
{\cal S}[\Bphi]= \int \left({\cal T}[\Bphi]-{\cal U}[\Bphi]\right)d^2xdt
\ee
where the kinetic energy density ${\cal T}[\Bphi]$ comprises all terms
containing time-derivatives of $\Bphi$. We consider the evolution of field
configurations, which at some initial time deviate from the global
equilibrium configuration (the 'true vacuum') within some finite
spatial region $V$. The equations of motion
$\delta{\cal S}/\delta\Bphi=0$ which govern the classical evolution
of the initial non-equilibrium configuration describe both, the
approach towards the vacuum configuration in the interior of that
spatial region and the propagation of outgoing (distorted) waves into
the (initially undisturbed) surrounding vacuum. Although, of course,
the classical equations of motion conserve the total energy
$E=\int \left({\cal T}+{\cal U} \right)d^2x$, the outgoing waves carry
away energy from the interior of the spatial region in which 
the total energy $E=T+U$  initially was located.

In quantum field theory the quantization of the propagating
fluctuations describes multiparticle production of pions emitted
from the relaxing field configurations. 
In a separation
\be
\label{sep}
\Bphi(\Bx,t)=\Bphi_{cl}(\Bx,t)+\delta\Bphi(\Bx,t)
\ee
where $\Bphi_{cl}$ comprises all of the large amplitude motion
of $\Bphi$, and $\delta\Bphi$ contains only small fluctuations around
$\Bphi_{cl}$, expansion of ${\cal S}[\Bphi]$ to second order
in $\delta\Bphi$
$$
{\cal S}[\Bphi]  =  {\cal S}[\Bphi_{cl}] 
 +  \int \frac{\delta{\cal S}}{\delta\Bphi(\Bx',t')}|_{[\Bphi_{cl}]}
\delta\Bphi(\Bx',t')d^2x'dt' 
$$
\be
\label{exp}
 +  \int\int \delta\Bphi(\Bx',t')\frac{\delta^2{\cal S}}{\delta\Bphi(\Bx',t')
\delta\Bphi(\Bx'',t'')}|_{[\Bphi_{cl}]}
\delta\Bphi(\Bx'',t'')d^2x'dt'd^2x''dt'' ~~+{\cal O}(\delta\Phi^3)
\ee
and variation with respect to $\delta\Bphi$ provides the classical
expression for the fluctuating part 
\be
\label{fluc}
\delta\Bphi(\Bx,t)=\delta\Bphi^{(0)}(\Bx,t) 
+\int {\cal G}(\Bx,t,\Bx',t')\frac{\delta{\cal
S}}{\delta\Bphi(\Bx',t')}|_{[\Bphi_{cl}]} d^2x'dt'.
\ee
The homogeneous part $\delta\Bphi^{(0)}$, after quantization,  represents
the scattering of field quanta off the classical configuration
$\Bphi_{cl}$ (i.e., these are on-shell distorted waves).
The Green's function $ {\cal G}(\Bx,t,\Bx',t')$ of the operator
$\delta^2{\cal S}/(\delta\Bphi \delta\Bphi)|_{[\Bphi_{cl}]}$
relates emitted (i.e on-shell) radiation to the source term 
$\delta{\cal S}/\delta\Bphi|_{[\Bphi_{cl}]}$.
There is, however, no unique way
to separate the propagating fluctuations $\delta\Bphi$ from a
more or less smoothly evolving classical configuration $\Bphi_{cl}$
because both, $\Bphi_{cl}$ and $\delta\Bphi$, are integral parts of one
and the same evolving order-parameter field $\Bphi$. Conclusions drawn
from one part only are subject to the arbitrariness of the chosen
separation.
In any case, in an equation of motion
that separately describes the evolution of $\Bphi_{cl}$
we expect a dissipative term to account for the loss of energy through
the emitted radiation:
\be
\label{deom}
\frac{1}{\tau}\dot{\Bphi}_{cl}=\delta {\cal S}/\delta\Bphi|_{[\Bphi_{cl}]}.
\ee
For sufficiently small values of the relaxation constant $\tau$
the dissipative  term dominates the time evolution of $\Bphi_{cl}$, 
propagating parts get damped away and we can
replace (\ref{deom}) by the Time-Dependent Ginzburg-Landau (TDGL) equation 
\be
\label{TDGL}
\frac{1}{\tau}\dot{\Bphi}_{cl}=-\delta U/\delta\Bphi|_{[\Bphi_{cl}]}.
\ee
The potential energy functional $U$ contains no time derivatives of
$\Bphi$, therefore $\Bphi_{cl}$ has no propagating parts, i.e. it does
not pick up field momentum, and it provides a very slowly moving
adiabatically evolving classical background field. 

Back-reaction of the fluctuations on the
moving classical field can re-enter in different ways: as loop
corrections they can lead to renormalization of the 
coupling constants in the functional $U$ which thus may become 
time- or temperature-dependent and
drive the spontaneous symmetry breaking; an appropriate noise
term to be added on the 
right-hand side of (\ref{TDGL}) accounts for stochastic interactions
of the fluctuations with $\Bphi_{cl}$ at finite temperature. 
Ideally, through that procedure one hopes to achieve a separation
(\ref{sep}) such that the fluctuating parts $\delta\Bphi$ contain no
exponentially increasing amplitudes during the relaxation
process.  

In the following we assume that the coupling constants which appear in 
the TDGL-equations have their renormalized physical  values with 
an appropriate time- or temperature dependence and we will restrict
the discussion to the consequences of those equations.
Evolutions via (\ref{TDGL}) of individual initial
$\Bphi_{cl}$ configurations  into 
final quasi-stable configurations are denoted as 'events', and we shall
consider ensembles of events sampled with different initial configurations.
The individual events show a variety of physically interesting
features, textures, walls, disoriented domains, growth laws, which form
the main body of the subsequent discussions. The lack of second-order
time derivatives in (\ref{TDGL}) means that also localized stable
structures created in $\Bphi_{cl}$ will not propagate but 
only drift in response to their mutual interaction. This is as
expected because uniform motion of textures represents zero modes
which may still be contained in $\delta\Phi$. 
With (\ref{TDGL}) the rate of decrease of the energy $U_{cl}=U[\Bphi_{cl}]$
stored in the classical configuration $\Bphi_{cl}$ then is
\be
\label{Udot}
\dot{U}_{cl}(t)=-\frac{1}{\tau}\int |\dot{\Bphi}_{cl}(\Bx,t)|^2 d^2x.
\ee
If particle emission from the evolving source is the only mechanism for
energy nonconservation 
this rate of energy loss has to match the energy flow carried away by
the emitted radiation. This requirement
determines the relaxation constant $\tau$ in (\ref{TDGL}) which fixes
the time scale 
for the approach of $\Bphi_{cl}$ towards equilibrium. 
(Of course, external interference with
the system like Bjorken rescaling of the spatial metric may provide
other sources of energy nonconservation.)

In quantum field theory, for field fluctuations coupled to a
time-dependent classical source $s(\Bx,t)$, the probability density 
for the emission of $n$$_{\scriptsize \Bk}$ field quanta with momentum $\Bk$ 
and mass $m$ is given by
\be
n_{\scriptsize{\Bk}}=|s(\Bk,\omega_k)|^2/(2\omega_k)
\ee
where $s(\Bk,\omega_k)$ is the energy-momentum-space transform of the
source, expanded in terms of the complete set of distorted
scattering states with asymptotic momentum $\Bk$, and taken at the
on-shell energy $\omega=\omega_k=\sqrt{k^2+m^2}$. 

In our present case according to (\ref{fluc}) the time-dependent
source is given by 
$\delta{\cal S}/\delta\Bphi|_{[\Bphi_{cl}(\Bx,t)]}$. Consistent with
our assumptions about the slow motion of the classical field
$\Bphi_{cl}$ we again may approximate here $\delta{\cal S}/\delta\Bphi$ by
$-\delta{\cal U}/\delta\Bphi$. With $\Bphi_{cl}$ obeying the TDGL
equations (\ref{TDGL}) the source field then is just the time-derivative
of the classical field $\dot{\Bphi}_{cl}(\Bx,t)/\tau$ and we obtain for
the average number of particles emitted with momentum $\Bk$ and isospin
component $i$ (omitting the index $cl$)
\be
\label{nios}
n_{i,\scriptsize{\Bk}}=\frac{|\dot{\Bphi}_i(\Bk,\omega_k)/\tau|^2}{2\omega_k}. 
\ee

With particle emission as the only mechanism for energy nonconservation
the total energy $\Delta E=\int \omega_k \sum_i n$$_{i,{\scriptsize \Bk}}
$$d^2k$ carried 
away by radiation during the ordering process must balance the loss of 
energy stored in the classical field. With (\ref{Udot}) this requirement
leads to
\be
\label{econs}
\frac{1}{2}\int|\dot{\Bphi}_{cl}(\Bk,\omega_k)|^2 d^2k 
=\tau\int|\dot{\Bphi}_{cl}(\Bx,t)|^2 d^2x dt.
\ee
Note that on the left-hand side of this equation only on-shell
frequencies appear, while the right-hand side (when written in Fourier
space) contains an independent integral
over all Fourier components of the moving classical field. Of course,
together with 
(\ref{TDGL}), this relation is an implicit definition of the
relaxation constant $\tau$.

For the adiabatically evolving process we consider the time $t$ as
parameter such that we may define
\be
\label{epsi}
\epsilon_i(t)=\frac{1}{\tau}\int \dot{\Phi}_i(\Bx,t)^2 \: d^2x
=\frac{L^2}{\tau} \langle \dot{\Phi}_i^2(t) \rangle
\ee
as the energy carried away per time interval by particles emitted with
field orientation $i$. Here "$\langle~\rangle$" denotes the lattice
average over an $L\times L$ lattice. Due to the slow motion of the
source we expect 
the particles to be mainly emitted with low energies, i.e. with
low momenta $k \approx 0$ and $\omega_i \approx m_i$. In our present
context we choose the $O(3)$-symmetry to be spontaneously broken in 
3-direction (selected by the surrounding true vacuum boundary condition
or by small explicit symmetry breaking, or by a small bias in the initial
configuration). Then the i=1,2 ('isospin') components of the
$O(3)$ chiral field constitute two 'pionic' fluctuational fields 
$\delta \Phi_i = \pi_i$ with small mass $m_{1,2}=m_\pi$ due to 
explicit symmetry breaking, while the
$\sigma$-fluctuation $\delta \Phi_3$ acquires a 
large mass $m_\sigma$ due to the spontaneously broken symmetry.
So we expect the emission rate for pions with isospin
component $i=1,2$
\be
\label{ni}
n_i(t) =\frac{\epsilon_i(t)}{m_\pi}
\ee
During the early stages of the ordering evolution the spatial averages
$\langle  \dot{\Phi}_i^2(t) \rangle $ will be similar for both
isospin directions, but at late times with the formation of larger
disoriented domains they might differ appreciably. The relative
abundancies 
\be
\label{fi}
f_i(t)=\frac{n_i(t)}{n_1(t)+n_2(t)}
\ee
are free of unknown constants and could serve as DCC signature, if it
is possible to separate in each individual event the small number of
late time pions from the background of those produced during earlier
stages, from decaying $\sigma$'s, and from other sources.

\section{The model}
The features which are of physical interest in the chiral
phase transition for a cooling hadronic gas in the 3+1 dimensional
$O(4)$ model can be explored and demonstrated in the lower-dimensional
$O(3)$ model. It is defined in terms of the dimensionless $3$-component
field $\Bphi = \Phi \hphi$ with $\hphi \cdot \hphi =1$,
with the following lagrangian density in $2+1$ dimensions

\be\label{lag}
{\cal{L}} = \frac{1}{2} \partial_\mu \Bphi \partial^\mu \Bphi
- \frac{\lambda}{4 \ell^2}\left( \Phi^4-2 f^2 \Phi^2 +1\right)
- \lambda\ell^2 \; \rho_\mu \rho^\mu 
+ m_\pi^2 \Phi_3.  
\ee
In addition to the usual linear sigma-model 
with standard  $\Phi^4$ potential for the modulus field $\Phi$
this lagrangian contains  
a four-derivative current-current coupling
$\rho_\mu\rho^\mu$ for the conserved 
topological current 
\be\label{top}
\rho^\mu = \frac{1}{8 \pi} \epsilon^{\mu \nu \rho} \hphi \cdot
( \partial_\nu \hphi \times \partial_\rho \hphi ) ,
\ee
which satisfies $\partial_\mu \rho^\mu = 0$. In the present $O(3)$
model this nonlinear coupling comprises the four-derivative
'Skyrme'-term and the sixth-order current-current coupling of effective 
chiral $O(4)$-models.

The independent strengths of the
$\Phi^4$- and the Skyrme couplings are written 
in terms of one
common dimensionless parameter $\lambda$, and a length $\ell$ 
which may be
absorbed into the space-time coordinates. So, for $\lambda$ fixed, 
$\ell$ sets the scale for the spatial extent of localized static
solutions, in lattice implementations relative to the size of the basic
lattice unit cell. The static solutions have been investigated and
described in~\cite{Ho}. For continuous coordinates (or, on a lattice,
for $\ell \gg 1$ ) the total energy (apart from the explicitly
symmetry-breaking $m_\pi^2$ contribution) of such localized solutions
is independent of $\ell$ (as long as $\ell$ is small as compared to the
total lattice size $L$). In the following we put the lattice constant
to unity. 

If we consider the field $\Bphi$ as the basic order parameter to
reflect the chiral phase transition as function of the temperature $T$,
then the factor $f^2$ which 
multiplies the square of the field vector in (\ref{lag}) drives the
spontaneous symmetry breaking. As the temperature goes to zero
it is supposed to converge towards its vacuum value $f^2(T\to 0) \to 1$,
and it should become very small 
$f^2(T\to T_c) \ll 1$ as $T$ approaches the critical temperature $T_c$.
For very high temperature $f^2(T > T_c)$ is supposed to become negative.

A small symmetry-breaking term with strength $m_\pi^2$ has also
been added to the otherwise O(3)-symmetric
lagrangian~(\ref{lag}). This explicit symmetry breaking is applied in
$3$-direction of the internal space. If unequal zero it singles out the
3-component $\Phi_3$ of the $\Bphi$-field as that field
which aquires the nonvanishing condensate at $T=0$, and we will denote
this component as "$\sigma$"-field, while the $1$- and
$2$-components represent the "pionic" (Goldstone-) degrees of freedom
of the remaining unbroken symmetry.

It is convenient to have also for nonvanishing $m_\pi^2$ 
the minimum of the ($T=0$)-potential located at the vacuum field
configuration $\Bphi=(0,0,1)$,
such that the length $\Phi$ of the field vectors approaches unity for
$T=0$ also for nonzero symmetry breaking. This is achieved by subtracting
$m_\pi^2 \ell^2/\lambda$ from $f^2$ in the coefficient of the second
order term in the potential $V(\Phi)$ in (\ref{lag})
\be
\label{f2}
f^2=f^2(T)-\frac{m_\pi^2 \ell^2}{\lambda}.
\ee
This form shows directly
how the explicit symmetry breaking supplies the pionic Goldstone
bosons with mass $m_\pi$.
For $T < T_c$ the temperature dependence of $m_\pi^2$ can be ignored.
On the other hand, the mass of the $\sigma$-field is given by
\be
\label{sigmass}
m_\sigma^2=2\frac{\lambda}{\ell^2}f^2(T)+m_\pi^2
\ee
which is mainly determined by the temperature dependence of $f^2(T)$.

Evolution  of $\Bphi_{cl}$  proceeds through the
TDGL equations of motion as they result from the potential part of
(\ref{lag}), (i.e. also from the
current-current coupling only the density term $\rho_0^2$ is kept).
As no other time-derivative terms appear on the right-hand side,
the constant $\tau$ which multiplies the left-hand side of
(\ref{TDGL}) can be set to one, i.e. in the following the relaxation
time $\tau$ serves as time unit. Omitting for convenience the
index $'cl'$ we have
\be
\label{eom}
\frac{1}{\tau}\dot{\Phi_i} = \Delta\Phi_i
-\frac{\lambda}{\ell^2} \left( \Phi^2-f^2 \right) \Phi_i
-\lambda \ell^2\; \frac{\delta(\rho_0^2)}{\delta \Phi_i}
+ m_\pi^2 \delta_{i3}
+ \xi_i.
\ee
As far as the dissipative term originates from elimination of
fluctuational modes which simulate a heat bath for the evolving field
configuration, addition of fluctuational noise $\Bxi$ is required
with a strength dictated by the dissipation-fluctuation theorem. Such a
term has also been added in (\ref{eom}). 

In lattice implementation, we impose periodic boundary conditions 
on the field configurations. This implies compactification of
coordinate space to a torus $S^1\times S^1$. These boundary conditions
preserve the global $O(3)$ symmetry of the model (for vanishing $m_\pi^2$).
Having singled out the 3-direction as the $\sigma$-direction in which
the $T=0$ condensate is formed we can impose the stronger vacuum
boundary conditions $\Bphi=(0,0,1)$ for all points on the lattice
boundary throughout the whole evolution.
This implies compactification of coordinate space to
the two-sphere $S^2$. In both cases the winding density $\rho_0$
satisfies $\int \rho_0 d^2x =B$ with integer winding number $B$.
Of course, these vacuum boundary conditions represent another way of
explicit symmetry breaking which eventually drives the evolution of the
condensate into the 3-direction. It is, however, a very weak form which is
only active at the surface of the spatial region, unlike a nonvanishing
$m_\pi^2$-term which is locally active at every point in space. 

With the equations of motion (\ref{eom}) being purely first order in time
derivatives it is sufficient to 
specify for initial conditions at  $t=0$ only the field configurations
themselves. As the time-dependent fluctuations are not part of
$\Bphi_{cl}$ the high-temperature initial configuration is,
(apart from a small optional bias in 3-direction due to an explicit symmetry
breaker), $\Bphi_{cl}\equiv 0$ in the interior of the spatial region
in which the hot chiral field is located and takes the true vacuum
values on its boundary. The first step in the time evolution
is therefore governed by the stochastic force $\Bxi$ alone. But this
first step will immediately create a configuration $\Bphi_{cl}$ which is of
stochastic nature itself. Therefore, choosing 
each of the three cartesian field components at each interior point
of the lattice  from a random Gaussian deviate, and fixing $\Bphi_{cl}=
(0,0,1)$  on the boundary of the lattice, provides convenient initial
configurations  with initial spatial correlation length less than one
lattice unit. From these random configurations, those with the desired
total winding number $B$ are selected and propagated in time. 
During the evolution of each individual event the 
fluctuational noise $\Bxi$ at every time step and every lattice point
is picked from an appropriate Gaussian deviate. A large number of
events generated in this way constitute the statistical ensemble from which 
ensemble averages can be obtained at some point in time during the evolution.

For configurations where the orientation of the vector field $\Bphi$
shows strong local variations (as it happens near localized textures or
throughout random initial distributions)
the implementation of the winding density $\rho_0$ on the lattice
in the differential form (\ref{top}) is not sufficiently accurate to
add up to integer total winding number $B$. Therefore $\rho_0$ and its
functional derivatives have to be evaluated according to the
geometrical meaning of $\rho_0$ as surface area of spherical triangles
cut out on the two-sphere $S^2$  by the image of elementary triangles
in coordinate space. Details of this procedure have been discussed 
in~\cite{HoKl}. It allows to assign a value $\rho(i,j)$ of the 
winding density $\rho_0$ to each point $(i,j)$ of the square lattice 
(or, more precisely, to each lattice cell with lowerleft corner $(i,j)$)
such that the total winding number
\be
B=\sum_{i,j=0}^{L-1} \rho(i,j)
\ee
summed over the whole lattice is integer and initial configurations can
be selected with some desired integer value of $B$. We also define 
\be
\label{D}
D=\sum_{i,j=0}^{L-1} |\rho(i,j)|
\ee
by summing up the absolute values of the local winding densities.
Of course, for random or slowly varying smooth configurations
$D$ generally is not an integer. However, if a configuration describes
a distribution of localized textures (and antitextures) which are
sufficiently well separated 
from each other, then $D$ is close to an integer and counts the number
of these textures (plus antitextures). In that case we can define the
numbers $N_+,N_-$ 
of "particles" and "antiparticles" through
\be
B=N_+ - N_-~~~~~~~~~~~D=N_++N_-~~.
\ee
In the following we will briefly call $D$ the "particle number", even
if it is not integer and comprises textures and antitextures. 

In a lattice implementation the local update of $\Bphi(i,j)$ at some
timestep at some lattice point $(i,j)$ will occasionally lead to a
discrete jump in the total winding number $B$. 
For well-developed
localized structures this corresponds to unwinding textures or
antitextures independently, such that $B$ decreases or increases by one
or more units.
This will eventually happen even for implementations of the nonlinear
$O(3)$-model where the length $\Phi$ is constrained to unity and $B$ is
topologically conserved, because the topological arguments based on
continuity do not apply to the discrete lattice configurations. 
It can be prevented by an (optional) $B$ filter which rejects such a
$B$-violating local field update at that lattice point and timestep.
This eliminates all independent unwinding processes. Only 
simultaneous annihilation of texture and antitexture in the same time
step remains possible, and, as the update proceeds locally at each
lattice vertex it can happen only if texture and antitexture overlap.
This $B$-conserving evolution is characteristic for the nontrivial
topology of the nonlinear $O(3)$-model and in this way can be
implemented as an optional constraint also into the linear $O(3)$-model. 
In the continuum limit it implies that a vanishing modulus $\Phi=0$ of the
field vector is excluded.

\section{Roll-down and Domain growth after sudden quench}

\begin{figure}[h]
\centering
\includegraphics[ width=12cm,height=15cm,angle=-90]{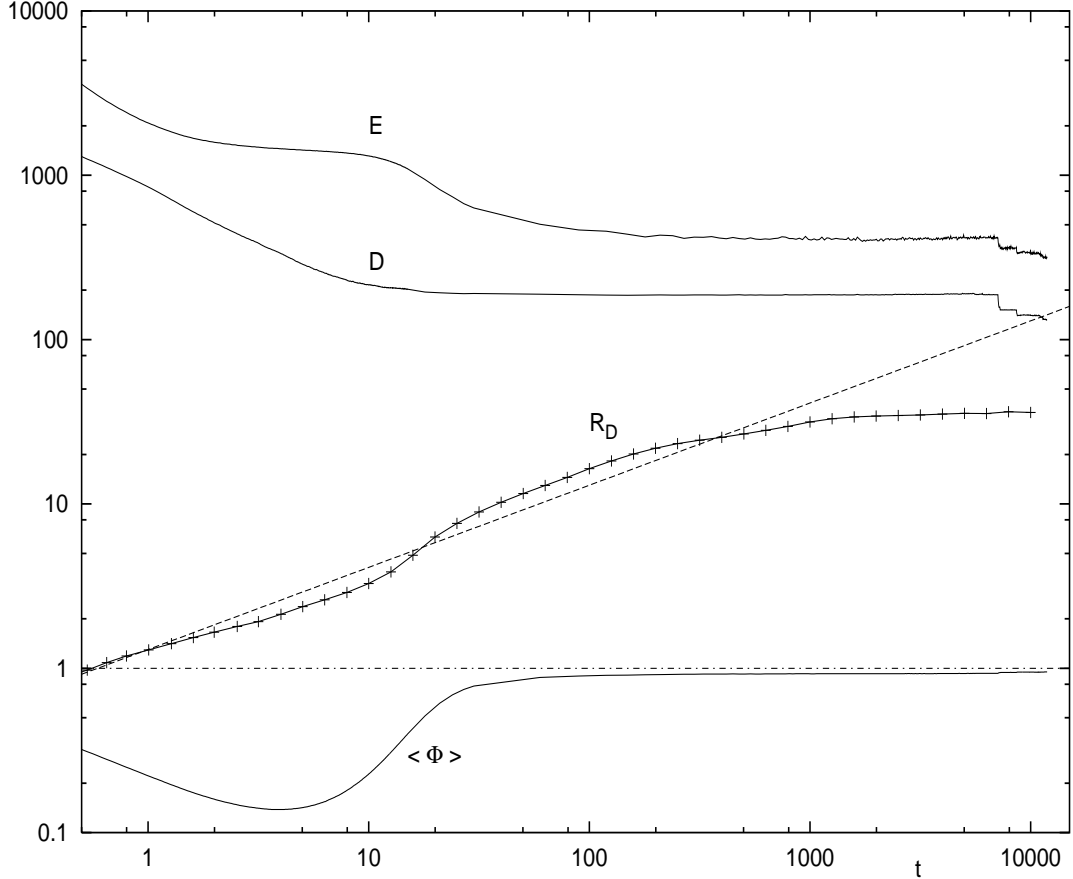}
\caption{Total energy $E$, particle number $D$, angular correlation
length $R_D$, 
and the average length of the field vector $\langle\Phi\rangle$ as functions of
time after a sudden quench (for $\ell=2,\lambda=1$). The total winding
number is $B=0$ by constraint  throughout the evolution. The dashed line
indicates the classical growth law $R_D\sim t^{0.5}$. }
\label{ell2}
\end{figure}

In order to get a feeling for a reasonable choice of the two essential
parameters $\ell$ and $\lambda$ in (\ref{lag}) we study at first the
simple case of a sudden quench without any symmetry breaking, neither
explicit nor via boundary condition, the so-called critical quench. 
That is, we put $m_\pi^2=0$ and 
impose only periodicity for otherwise arbitrary field vectors at the
boundary of a sufficiently large $L\times L$ lattice. The sudden quench
is realized by exposing the initially (at $t=0$) random configuration 
to the $T=0$-value of $f^2(T=0) \equiv 1$ for all times $t>0$. For this
zero-temperature evolution the noise term $\Bxi$ is also not present.
As an example fig.\ref{ell2} shows one individual evolution of a configuration
with fixed winding number $B=0$, on a lattice with $L=150$, with
$\ell=1$ and $\lambda=2$.

As significant observables we show the time-dependence of total
energy $E$, particle number $D$, field modulus $\langle\Phi\rangle$ averaged over the
whole lattice, and the angular correlation length $R_D$.
The latter is defined as the (interpolated) half-maximum distance $R_{1/2}$
of the equal-time angular correlation function
\be
\label{corr}
C(R)=\sum_{i,j=0}^{L}\sum_{k,l=0}^{L} \hphi(i,j)\cdot \hphi(k,l)
/\sum_{i,j=0}^{L}\sum_{k,l=0}^{L} 1
\ee 
(with the $k,l$ sum restricted such that 
the distance $r=\sqrt{(k-i)^2+(l-j)^2}$ between lattice vertices
$(i,j)$ and $(k,l)$ lies inside bins of unit size around
fixed positive integers $R$). Note that the definition (\ref{corr})
contains only the unit vectors $\hphi$, and not the full field
vectors $\Bphi$, i.e. $C(R)$ measures only the angular correlations in
a given configuration. As discussed in~\cite{HoKl} this definition of
$R_D$ serves well to provide an average radius of oriented domains.

Fig.\ref{ell2} follows the evolution over a very long time (up to $t\sim 10^4$
relaxation-time units) and clearly shows the different phases of 
the ordering process. Up to $t \sim 10$ the average $\langle\Phi\rangle$ stays 
(or becomes) small ($\sim 0.1$), while the particle number $D$ decreases
by almost one order of magnitude due to local angular reorientation,
also reflected in $R_D$ which increases approximately as
$t^{0.4}$. Once the oriented domains have reached sizes which cover
several lattice units the roll-down phase sets in where $\Phi$
approaches unity within these domains (while remaining small at their
boundaries where strong angular variations prevail). This roll-down
phase takes a few tens of time units and it is accompanied by a
significant energy reduction and a more 
rapid increase of $R_D$, such that by the end of this roll-down the
oriented domains cover areas with typical radii of 10 lattice units.
During the end of, and after this phase the isolated textures assume
their stable 
shapes with the appropriate bag and angular field profiles. This is
accompanied by further angular ordering of the topologically trivial field
configurations in which these textures are embedded, which proceeds by
moving boundaries between differently oriented domains. It leads to
further decrease of $E$ and increase of $R_D$, which follows very
nicely the classical $t^{0.5}$ Allen-Cahn law~\cite{AlCa}. This phase
lasts up to several hundreds of time units, without further reduction
of the particle number $D$. Due to the existence of the remaining stable
bags the average $\Phi$ does not reach the vacuum limit $\langle\Phi\rangle=1$ and
due to the floating angular textures the increase in $R_D$ begins to saturate
after $t\sim 10^3$. On a timescale which is still larger by another
order of magnitude the stable isolated textures may meet and annihilate
as can be observed in fig.\ref{ell2} near $t\sim 10^4$.

\begin{figure}[h]
\centering
\includegraphics[width=9cm,height=12cm,angle=-90]{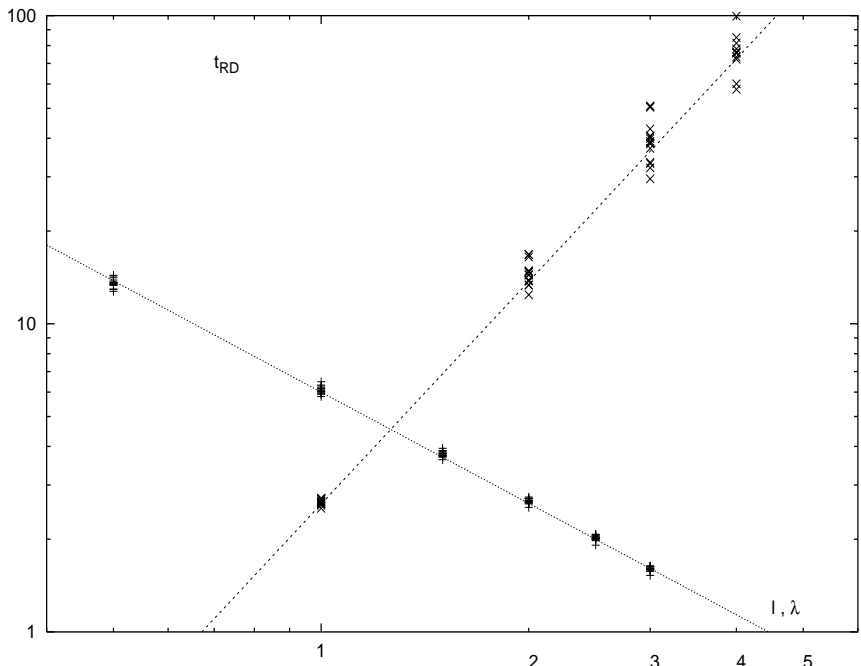}
\caption{ The roll-down time $t_{RD}$ for different values of
the scale $\ell$ (with coupling constant $\lambda=2$), and different
values of $\lambda$ (with scale $\ell=1$), respectively. For each set
of parameters 15 events ($B=0$) on a $120\times120$ grid are sampled.
The straight lines indicate the power laws $t_{RD}=2.6 \;\ell^{2.4}$
and $t_{RD}=6 \lambda^{-1.2}$, respectively.}
\label{tRDfig}
\end{figure}

Fig.\ref{ell2} is obtained for $B$ constrained to $B=0$ and for scale
$\ell=2$, such that resulting bag structures extend only over a few
lattice cells. The long-time behaviour
after the roll down phase is higly sensitive to the size of the 
developing stable textures, i.e. to the fraction of lattice space
occupied by the bags because, evidently, overlapping textures quickly combine
or annihilate. Their size is governed by $\ell$, and in the range up to
$\ell<5$ the structures created are reasonably isolated.
We shall see later that for the point of view of DCC signatures the
long-time behaviour after the roll-down is irrelevant. 
Defining the typical roll-down time $t_{RD}$ as that point in time
when the lattice averaged $\langle\Phi\rangle$ has reached a
value of 0.8, we 
plot in fig.\ref{tRDfig} $t_{RD}$ as function of $\ell$ and $\lambda$,
respectively, for 15 events each. For $B$ constrained to $B=0$ the
average values follow power 
laws $t_{RD} \sim \ell^{2.4}$ and $ t_{RD} \sim \lambda^{-1.2}$ with 
good accuracy. These numerical results show that the roll-down time
$t_{RD}$ scales as $(\ell^2/\lambda)^{1.2}$ which means that it is
essentially determined by the potential term $\left( \Phi^2-1\right)^2$
alone. It is almost unaffected by the fourth-order derivative term
which scales with the product $(\ell^2\lambda)$, while the second-order
gradient terms in $U$ cause only a mild increase of the scaling exponent
to the observed value of 1.2.

We interpret the stable textures as baryons floating in the effective
meson pion field. Fixing their spatial extent in terms of lattice units
through the choice of $\ell$ provides the physical scale for the
lattice constant. We shall typically choose $\ell=2$ such that baryons
carrying one unit of winding number (=baryon number) acquire a radius
for their topological density of about two lattice constants.
With a typical nucleon radius of 0.6 - 1.0 fm, for $\ell=2$, the
lattice constant then corresponds to $\sim 0.3 - 
0.5$ fm. For an accurate description of the texture
and bag profiles of individual baryons such a small value $\ell=2$
is not really sufficient, as has been discussed in detail in the
investigation of static solutions in~\cite{HoKl}. 
But within the present context this aspect is not essential.
For physical vacuum boundary conditions this choice also suggests
the typical lattice size $L$, which should be sufficiently large to
cover the spatial extent of the high-temperature area after a typical
hadronic (heavy-ion) collision. So values of $20<L<100$ seem
appropriate, and they are technically easy to handle. 

Through (\ref{sigmass}) the value of $\lambda$  is directly tied to the 
$\sigma$-mass. For $\ell=2$ the choice $\lambda=2$ leads to
$m_\sigma^2=1$ (in inverse lattice units), which appears quite
reasonable, being of the order of a typical baryon mass.
In the discussion in~\cite{HoKl} we have seen that for
$\lambda=2$ the stable textures are confined into the interior of
well-developed bag structures which we consider an essential feature
of physical baryons. For all these reasons we shall in the following
present results mostly for $\ell=2$ and $\lambda=2$.

In order to obtain the roll-down times for that case
we extract from fig.\ref{tRDfig} the values of $t_{RD}$ 
for this two-dimensional $O(3)$-model
\be
\label{tRD}
t_{RD}\approx 6 \;\left( \frac{\ell^2}{\lambda} \right)^{1.2}.
\ee
For  $\ell=2$, $\lambda=2$ this is about 14 relaxation time units.

To guide the eye for the increase of the correlation length $R_D$ 
we have included in fig.\ref{ell2} the power law $R=1.3 \; t^{0.5}$ as 
straight dashed line. Although there are typical deviations prior and
during the roll-down phase it is remarkable that this classical scaling
expectation serves well to describe the growth of oriented domains
for this rather complicated interplay of forming and stabilizing
extended textures and aligning domains. Apparently, the use of
this growth law for estimating domain sizes seems justified throughout
the whole time period relevant for DCC signatures.

\begin{figure}[h]
\centering
\includegraphics[width=9cm,height=12cm,angle=-90]{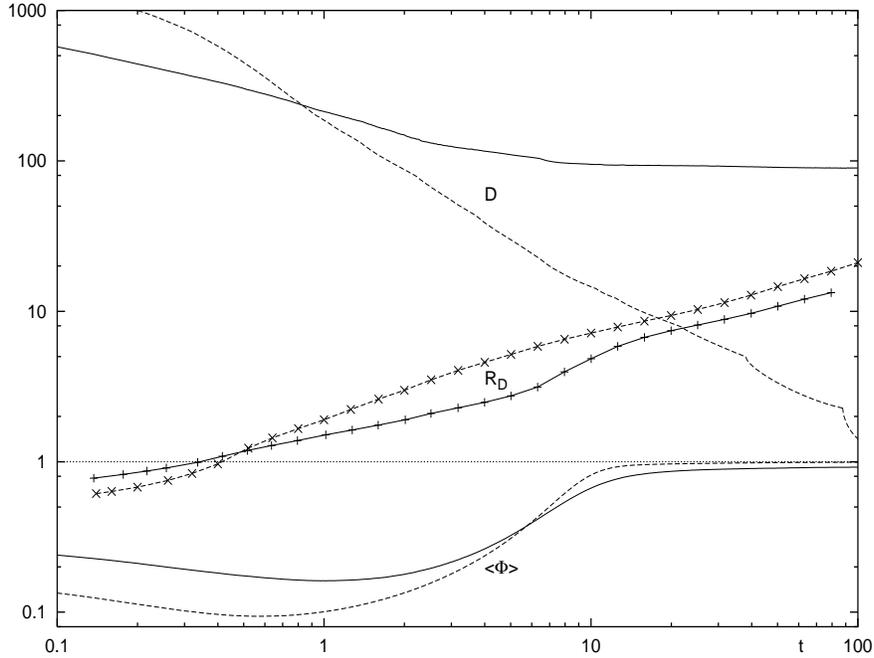}
\caption{Comparison of particle numbers $D$, angular correlation lengths
$R_D$, and the average bag fields $\langle\Phi\rangle$
as functions of time $t$ after a sudden quench, 
for evolutions under the full model with constraint $B=0$ (full lines),
and the pure linear $\sigma$-model (no fourth-order term, no constraint
on $B$ (dashed lines); (100$\times$100 grid, $\ell=2,\lambda=2$). }
\label{linsig}
\end{figure}

It is interesting to compare these results with the plain linear
$\sigma$-model. For that purpose we omit in (\ref{lag}) the
current-current coupling and follow the evolution without putting
a constraint on the winding number $B$. Fig.\ref{linsig} shows two
events evolving from identical initial configurations on a
100$\times$100 grid with vacuum boundary conditions. After about
200 relaxation time units the unconstrained linear sigma model has lost
all particles and antiparticles ($D \to 0$), and the configuration 
tends towards global alignment in 3-direction. 
On the other hand, the full model
with fourth-order Skyrme term included and winding number constrained
to $B=0$, finally stabilizes solitons with $D$ of the
order of 100 particles plus antiparticles. 
Comparing the growth rates of the angular correlation lengths $R_D$ 
for both evolutions it may be seen that there is no dramatic
difference in the overall growth exponent.
Still, it is evident that especially during the roll-down phase
the emerging extended stabilizing textures do present an obstacle
against the free and unconstrained ordering process such that the
average size of the ordered domains in the full model is noticeably 
smaller than in the plain linear $\sigma$-model evolution.

At this point we may add a few remarks concerning the relaxation
constant $\tau$ which in the TDGL-equation (\ref{eom}) we have simply
used as a time unit, because no other propagating terms are present.
Of course, we could use (\ref{deom}) instead of (\ref{TDGL}) for 
simulating evolutions of the classical field. Keeping the propagation
velocity at $c=1$ the time unit then is set by the lattice constant.
A trivial consequence is that all features which reflect the relaxation,
like growth rates or saturation times, are scaled with powers of $\tau$. 
E.g., for sufficiently small $\tau$ the roll-down time $t_{RD}$ scales
like $1/\tau$. 
It turns out that for $\tau \sim 0.1$ the evolutions resulting from
(\ref{deom}) and (\ref{TDGL}) essentially coincide. For $\tau \sim 1$ the
propagating waves start showing up in the field configurations and 
cause oscillatory features in (lattice-) averaged quantities without
strongly affecting the late-time evolution. For $\tau > 2$ oscillations
dominate and due to the nonlinear terms stabilization becomes
difficult. So, to be on the safe side for use of the TDGL equation
(\ref{TDGL}), $\tau$ should be less than 1.
Looking once more at fig.\ref{linsig}, $t_{RD}\sim 10$ (in units of
$\tau$) implies that for $\tau \sim 0.1$ the 
roll-down is completed after travelling waves have passed about 100
lattice units.

\section{Cooling times} 

The assumption of a sudden quench which underlies the ordering
features shown in the previous chapter represents that limit for the
actual cooling process where the cooling time $\tau_c$ (during which the
temperature drops from $T/T_c \sim 1$ to $T=0$) is much smaller than the 
relaxation time which governs the evolution of the
field configurations. This instantaneous cooling, 
where the hot initial configuration for $t>0$ is immediately exposed to the
zero-temperature $T=0$ effective potential defines the most off-equilibrium
scenario which we can study within this model. 
To weaken this rather drastic assumption we replace it with an
appropriate time dependence of the temperature $T(t)$. Then also
the temperature dependence  $f^2(T)$ in (\ref{f2}) has to be specified.

The function $f^2(T)$ in principle is determined from loop
contributions and their renormalization group resummation in 
finite-temperature field theory. Perturbative evaluations are
restricted to the vicinity of $T=0$; the complete form 
of $f^2(T)$ depends on the specific model and is not accurately known.  
Its essential features, however, are a slow decrease from its $T=0$
value $f^2=1$ for $T>0$, followed by a steeper decrease $f^2 \to 0$ as
$T$ approaches $T_c$. For the simulations these essentials are 
sufficient; the specific form (generally in form of logarithms) is
not important. A very convenient ansatz which comprises these basic
features in the region $0<T<T_c$ is
\be
\label{f2T}
f^2(T)= (1-(T/T_c)^\alpha))^2
\ee
with $\alpha \sim 4$,  and $f^2(T)=0$ for $T\ge T_c$.

A slow cooling scenario can be realized by externally imposing the
dependence $T(t)$ of temperature on time. For a linear quench we take
\be
\label{quench}
T(t)/T_c= 1-t/\tau_c
\ee
with quench time $\tau_c$, and $T=0$ for $t>\tau_c$.
A crucial feature of the ordering process is that the early
phase prior to the onset of the roll-down is dominated by angular
aligning driven by the gradient terms in the $\sigma$-model part of
(\ref{lag}) while the effective potential for $\Phi$ is of only minor
importance. As a consequence, the angular ordering up to the onset of
the roll-down proceeds quite independently of the actual cooling rate,
with $\langle\Phi\rangle\ll 1$ in any case. This is fortunate 
because it renders knowledge of the precise form of $T(t)$ and $f^2(T)$
unimportant and we can safely adopt convenient ansaetze like
(\ref{f2T}) and (\ref{quench}).

\begin{figure}[h]
\centering
\includegraphics[width=11cm,height=15cm, angle=-90]{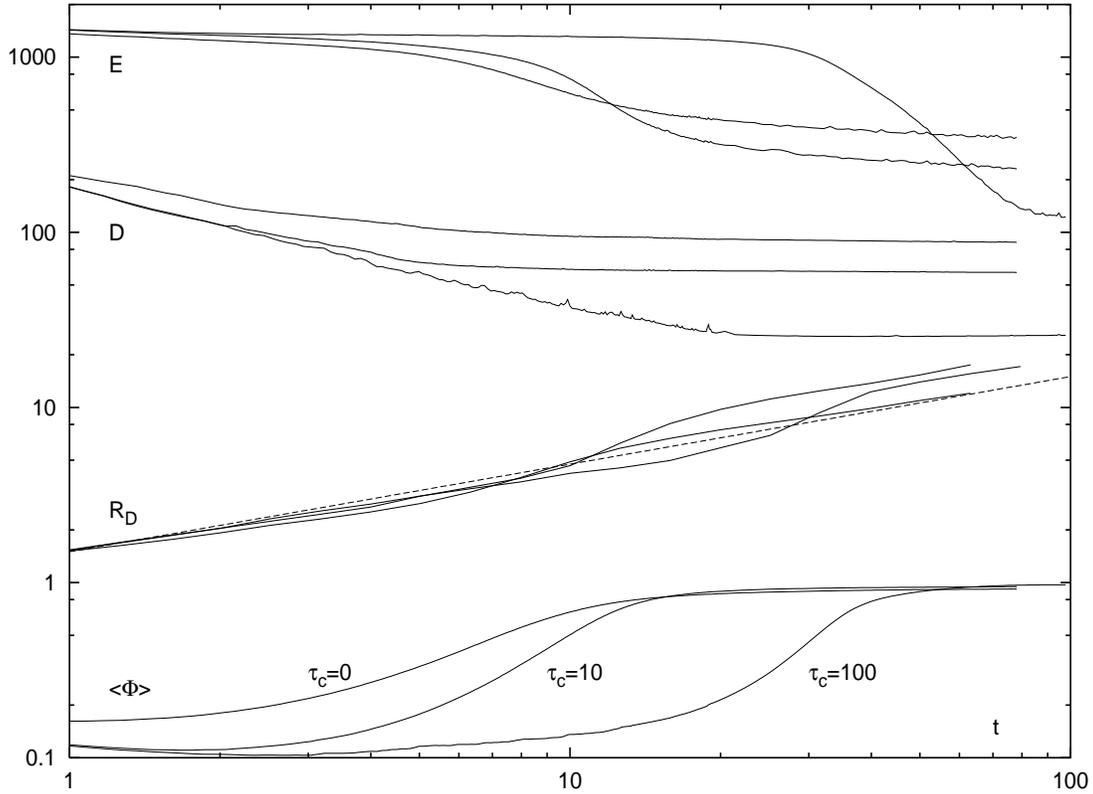}
\caption{Energies $E$, particle numbers $D$, angular correlation lengths
$R_D$, and average bag fields $\langle\Phi\rangle$ for three events
(100$\times$100 grid, $\lambda=2, \ell=2$) with linear quench (24) and
three different cooling times: $\tau_c=0$ (sudden quench), $\tau_c=10$
(intermediate), $\tau_c=100$ (very slow quench). The dashed line
indicates the growth law $R_D \sim t^{0.5}$. }
\label{tau}
\end{figure}

For a very slow quench $\tau_c \gg \tau$, after this first angular
aligning phase, the roll-down then only proceeds up to the
momentaneous value of $f^2$, $\langle\Phi\rangle^2 \le f^2(T(t))$.
Specifically, the stable textures are being created with profiles which
correspond to an instantaneous scale
\be
\label{ell}
\tilde \ell(t)= \ell f^{-1}(T(t)) 
\ee
as can be seen from rescaling the modulus of the order
parameter to $\tilde \Phi =\Phi f^{-1}(T(t))$ in (\ref{lag}). With
$f^2(T(t)) < 1 $ the nascent stable objects in the slow quench are much
larger than for the sudden quench, i.e. fewer of them survive as separate
isolated textures. Subsequently, further ordering proceeds essentially as
equilibrium evolution where the stable objects shrink
according to $\tilde \ell(t)$ in (\ref{ell}) until they finally assume their
$T=0$ size with accompanying rearrangement of surrounding aligned domains.

As intermediate case we consider evolutions where the cooling
time is comparable with the roll-down time $t_{RD}$ in (\ref{tRD}),
conveniently again as 
linear quench with $\tau_c \sim 10$ for $\ell=2$ and $\lambda=2$.
In the very early phase the evolutions closely follow the slow
quench with very small $\langle\Phi\rangle$ and steeper decrease of the
particle number $D$ as compared to the sudden quench, but as $t$
approaches $\tau_c$ the roll-down proceeds
rapidly towards the sudden quench value $\langle\Phi\rangle \approx 1$, with
corresponding pronounced loss in energy $E$ and typical increase of
$R_D$. For $t>t_{RD}$ the number of textures with $T=0$
profiles stabilizes at values intermediate between sudden and very
slow quench and the angular ordering of aligned vacuum domains proceeds
along the classical $t^{0.5}$-law.

\begin{figure}[h]
\begin{center}
\includegraphics[width =5cm,height=5cm,angle=-90]{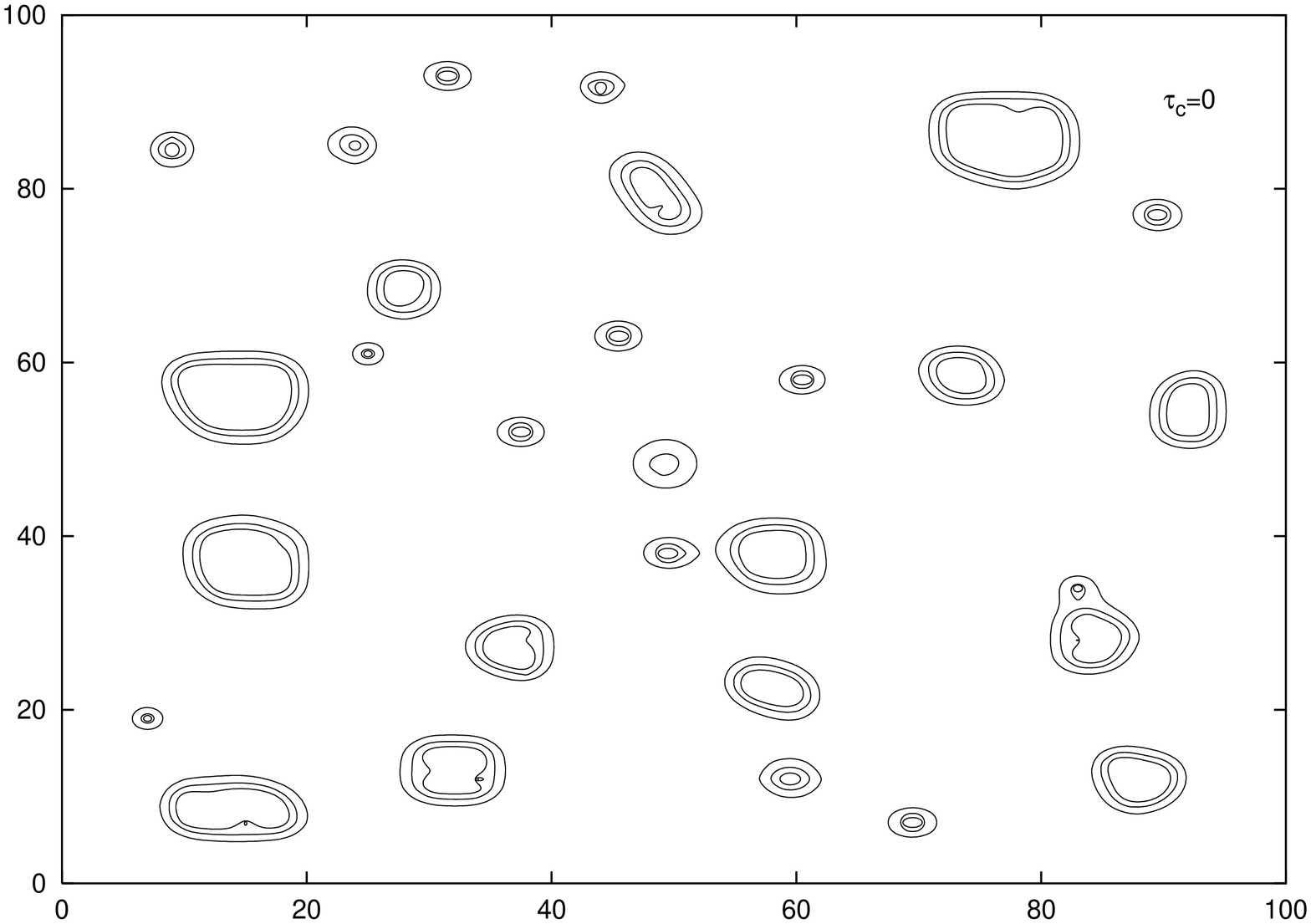}
\includegraphics[width =5cm,height=5cm,angle=-90]{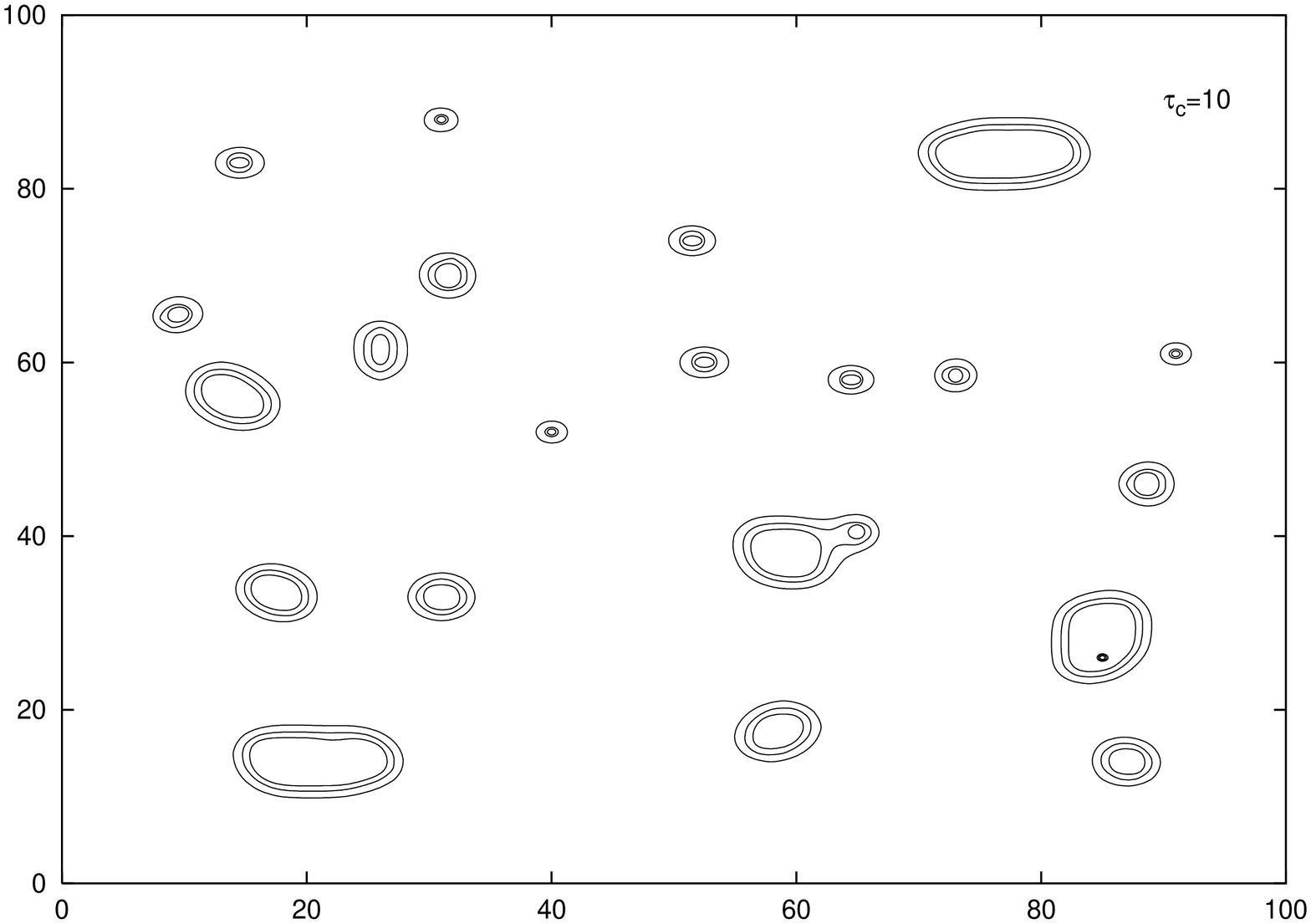}
\includegraphics[width =5cm,height=5cm,angle=-90]{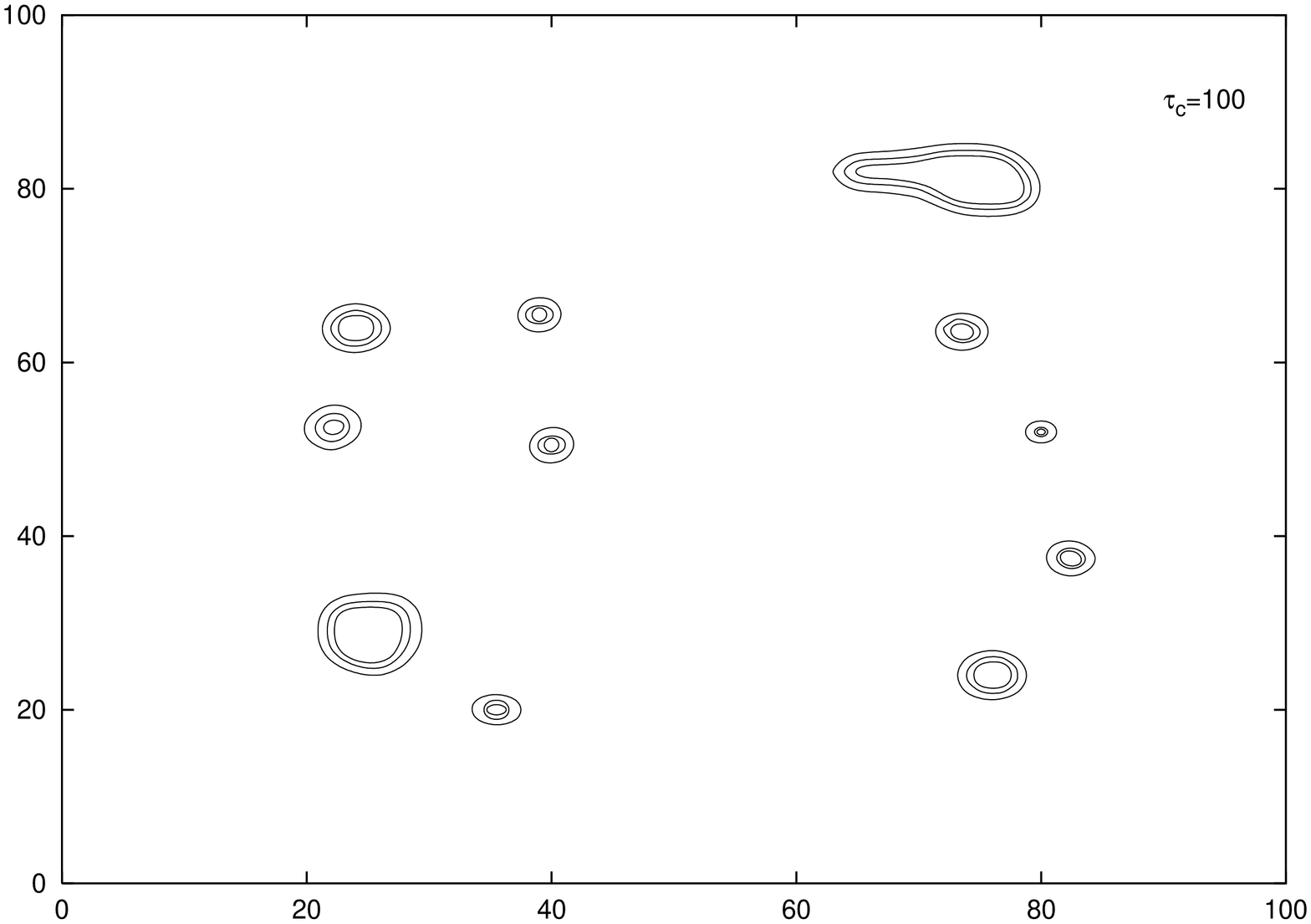}
\end{center}
\caption{ Bag contours (drawn at $\Phi=0.8,0.6,0.4$ from outside to
inside) of isolated baryons remaining at time $t=100$, for the 3 events shown
in fig.\ref{tau} which differ by their cooling times $\tau_c=0,10,100$. 
By the time $t=100$ the temperature has arrived at $T=0$ in all 3 cases,
and the baryons approach their $T=0$ sizes and quasistable shapes.
The smallest ones carry one unit of winding number. The total winding
number is zero in all three cases. The remaining particle numbers (as can
be seen from fig.\ref{tau}) are $D=95, 65, 24$ for the cooling times
$\tau_c = 0, 10, 100$, respectively. }
\label{ctr}
\end{figure}

Fig.\ref{tau} shows three typical evolutions with $\ell=2$ and
$\lambda=2$ and total winding number constrained to $B=0$,
starting from the same initial configuration, for $\tau_c = 100$ (very
slow), $\tau_c = 10$ (intermediate), and $\tau_c = 0$
 (sudden) quench. Fig.\ref{ctr} presents bag contours at
$\Phi=0.4,0.6,0.8$ of the stable textures with varying individual
winding numbers (summing up to $B=0$) which have been established after
$t\sim 100$ for 
these three different cooling rates. It is quite evident from fig.\ref{tau}
that the rate of increase of the correlation lengths $R_D$ before and
after the roll-down ($t\sim 10-20$) is unaffected by the
cooling rate. Differences in the angular ordering, however, show up in
the evolving particle numbers which reflect the changing effective potential
through the size of nascent textures.

For these slow quenches which proceed with finite cooling rate 
evolutions are
subject to fluctuational noise given by the last term in (\ref{eom}).
Its variance depends on the physical nature of the dissipative $\dot{\Bphi}$
term in (\ref{eom}), i.e., to which extent it originates 
purely from particle emission, or from
kinematical expansion of the spatial volume, or from a heat bath of
eliminated high-frequency modes. In any case, its strength according to the
fluctuation-dissipation theorem decreases linearly with temperature
$T$. Its general influence on the evolving configurations is very
transparent: for finite $T$ it will tend to disturb the ordering process at
each time step. This results in a slight decrease of the slopes of
those observables which reflect the ordering, specifically the
correlation length $R_D$, the average $\langle\Phi\rangle$, the particle
numbers $D$, 
the total energy $E$. On the other hand, fluctuations support
(Brownian) motion of nascent textures which combine more easily into
larger structures with increased individual winding numbers. So, 
evolutions subject to noise finally produce increased values of $D$
distributed into larger clusters of positive or negative winding
number. Naturally, only very slow quenches are noticeably affected
by the noise. For definiteness, the evolutions shown in figs.\ref{tau} are
obtained from (\ref{eom}) with the last term $\Bxi$
omitted completely. Effectively, inclusion of noise produces evolutions
which resemble those for quenches with slightly reduced quench time.

As we have seen, the physically most relevant characteristic of
the ordering evolution is the roll-down time where the bag field
$\Phi$ approaches within aligned (but still mostly disoriented) domains 
the momentaneous value of $f(T)$ and isolated
textures are being established near their borders.
This important feature is most
clearly brought forth for $\tau_c \to 0$ therefore in the following we
preferably will consider ordering processes after instantaneous cooling.

\section{Explicit symmetry breaking}

Explicit symmetry breaking of the global $O(3)$-symmetry is of crucial
importance for the ordering process. Its weakest form is imposed through
boundary conditions where a specific vacuum orientation, e.g.
$\Bphi=(0,0,1)$, along the lattice boundary is enforced. For 
physical processes this, apparently, is also a most natural condition.
With increasing time after a sudden quench the influence of this
surrounding vacuum will grow deeper and deeper into the randomly
ordering interior and eventually reorient aligned domains along the
chosen vacuum direction. Clearly, the strength of this aligning effect
depends on the finite size $L$ of the lattice. For $L$ of the order of
100 it does not change the essential features of the evolution.
There is only one ordering parameter, the absolute value of the
averaged 'direction' 
\be
\langle\hphi\rangle =   \sum_{i,j=0}^{L-1} \hphi(i,j) /L^2.
\ee
which is sensitively affected.
For arbitrary periodic boundary conditions this observable remains very
small with 
erratic variations for times long after the roll-down, while for 
the symmetry-breaking vacuum boundary conditions, it continually grows
in accordance with the growth of the angular 
correlation length. It is saturated by the averaged 3-component
$\langle\hphi_3\rangle$ which carries all of the total
$\langle\hphi\rangle$ strength.
So, as expected, the oriented domains growing in the vicinity of the
boundary contribute coherently to the magnetization and align in vacuum
direction.

\begin{figure}[h]
\centering
\includegraphics[width=9cm,height=12cm,angle=-90]{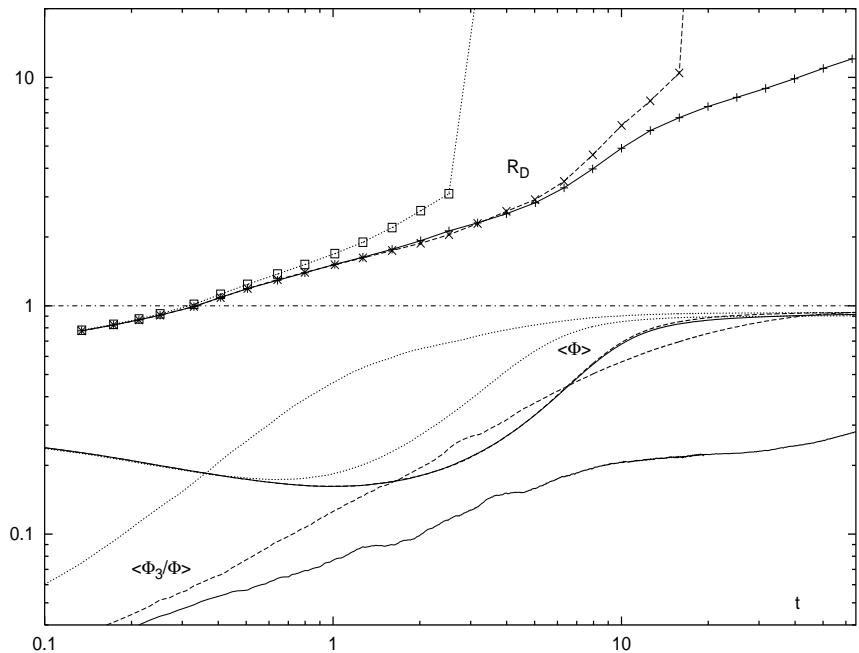}
\caption{ Angular correlation lengths $R_D$, average bag fields
$\langle\Phi\rangle$, and the averaged 3-component of the field unit
vector $\langle\hphi_3\rangle$, 
for three events (100$\times$100 grid, $\lambda=2, \ell=2$) 
with different strengths of explicit symmetry breaking, $m_\pi^2=0$
(full lines), $m_\pi^2=0.01$ (dashed lines), $m_\pi^2=0.1$ (dotted
lines). (Note that the average bag fields $\langle\Phi\rangle$ for
$m_\pi^2=0$ and $m_\pi^2=0.01$ almost coincide.) }
\label{csb}
\end{figure}

Explicit symmetry breaking through a nonvanishing $m_\pi^2$-term in
($\ref{lag}$) exerts much stronger influence on the ordering process
than imposing symmetry-breaking boundary conditions only. We compare in
fig.\ref{csb} 
evolutions under vacuum boundary conditions for values of $m_\pi^2=0.01$,
$0.1$, and $m_\pi^2=0$. This appears as a reasonable range for pions
with mass of about 10\% of the $\sigma$-mass. 
For $m_\pi^2=0.01$ the 
roll-down of $\langle\Phi\rangle$ is almost unaffected as compared to
$m_\pi^2=0$, 
while $m_\pi^2=0.1$ accelerates the roll-down by a factor of two (with
a correspondingly larger number of stable textures created).

Both evolutions 
with $m_\pi^2 \neq 0$ show the exponential approach of
$\langle\hphi_3 \rangle$ to saturation expected for the 
off-critical quench, in contrast to the unbroken case. 
(For infinite systems the
exponential approach to saturation is governed by $\exp(-2 m_\pi^2
t)$~\cite{BrayII}). The saturation value is smaller than one
due to the presence of stable textures.  In both cases
$\langle\hphi\rangle$ is completely saturated by $\langle\hphi_3\rangle$. 
It may also be noted from fig.\ref{csb} that for $m_\pi^2=0.1$ saturation
of $\langle\hphi_3\rangle$ is achieved long before the roll-down of
$\langle\Phi\rangle$ is 
completed, while for $m_\pi^2 = 0.01$ the order is reversed.

It should be noted that the angular
correlation function $C(R)$ in the off-critical system approaches
$\langle\hphi\rangle^2$ for $R \gg 1$. This shows up as an abrupt
increase of the half-maximum distance $R_D$, as soon as the average
$\langle\hphi\rangle $ has grown 
beyond $\sqrt{0.5} \approx 70\%$ of its saturation value. Before that 
sudden increase, $R_D$ already starts to deviate from the growth law
of the critical system, because it reflects also the increasing global
alignment in 3-direction. So, in the off-critical system, it appears
more appropriate to consider only the angular correlations of the
transverse ($i=1,2$) components as a measure for the size of
disoriented domains.

Additional breaking of the $O(3)$-symmetry occurs through an initial bias
in the starting configurations. It originates from the asymmetry 
introduced into the initial $T > T_c$ ensemble through the $m_\pi^2 \Phi_3$
symmetry-breaking term in (\ref{lag}). It will not alter the exponent
of the exponential approach to saturation, but it will reduce the
factor in front of the exponential. This causes another considerable
reduction of saturation times.

\section{Pion emission as DCC signature }

As we discussed in sect.2 signatures of the evolving disoriented
domains may be carried in the fractions $f(\pi_i)$ of pion multiplicities
of kind $i$ emitted at late time during the ordering process.
We will investigate these signatures for an ensemble of events
evolving after a sudden quench. 
We compare results for the plain linear $\sigma$-model,
(i.e. omitting the four-derivative current-current coupling in
(\ref{lag}), without constraint on winding number $B$), with the 
corresponding evolutions for the full lagrangian (\ref{lag}), 
with $B$ constrained to $B=0$, which lead to stabilized solitons
floating in the surrounding vacuum. For the parameters in the energy
functional we choose $\ell=2$ and $\lambda=2$, such that the emerging
solitons develop deep bags and extend over several lattice cells.
With this choice the mass of the (T=0) $\sigma$-fluctuations according to
(\ref{sigmass}) is $m_\sigma=1$.

\begin{figure}[h]
\begin{center}
\includegraphics[width =7cm,height=8cm,angle=-90]{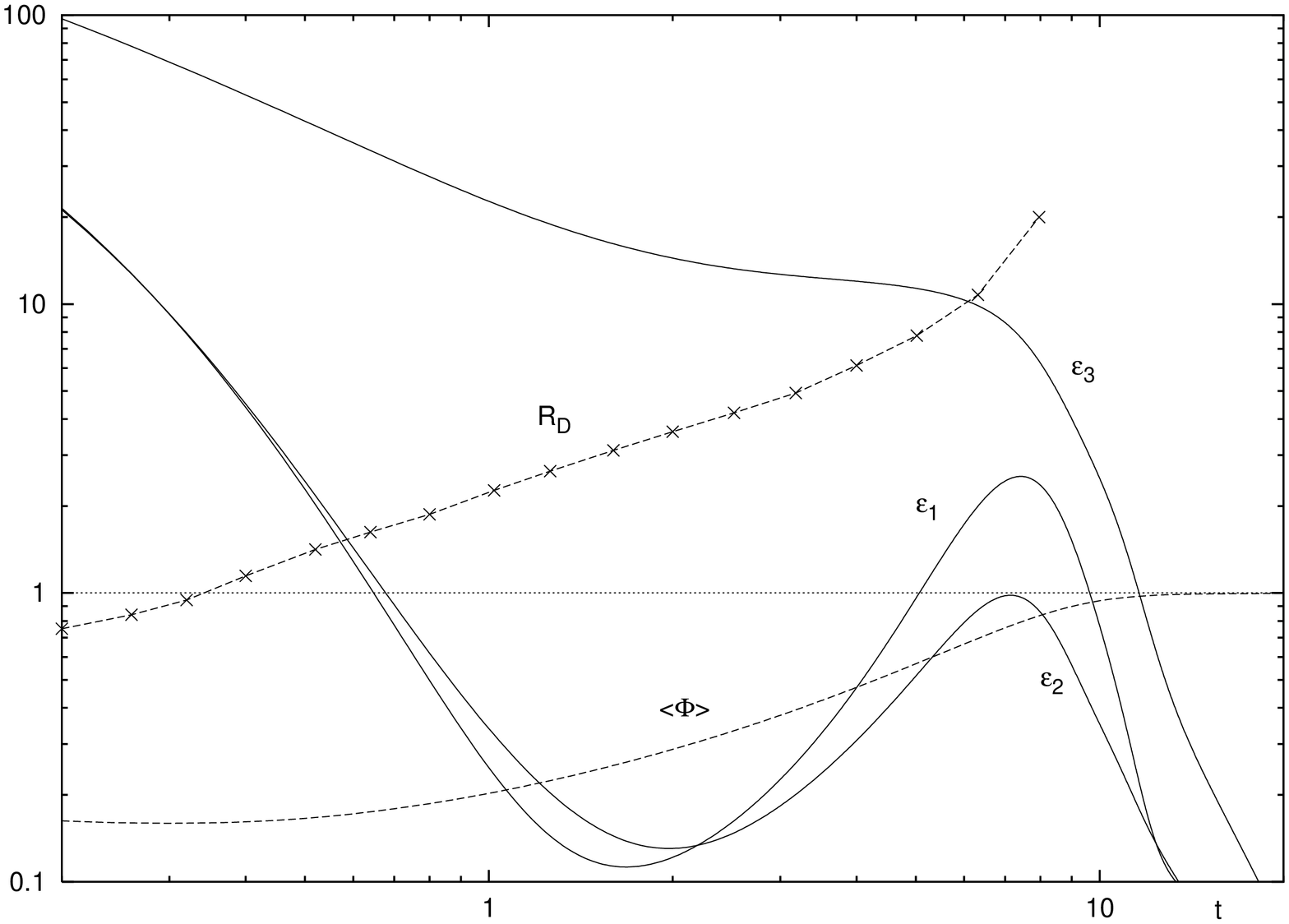}
\includegraphics[width =7cm,height=8cm,angle=-90]{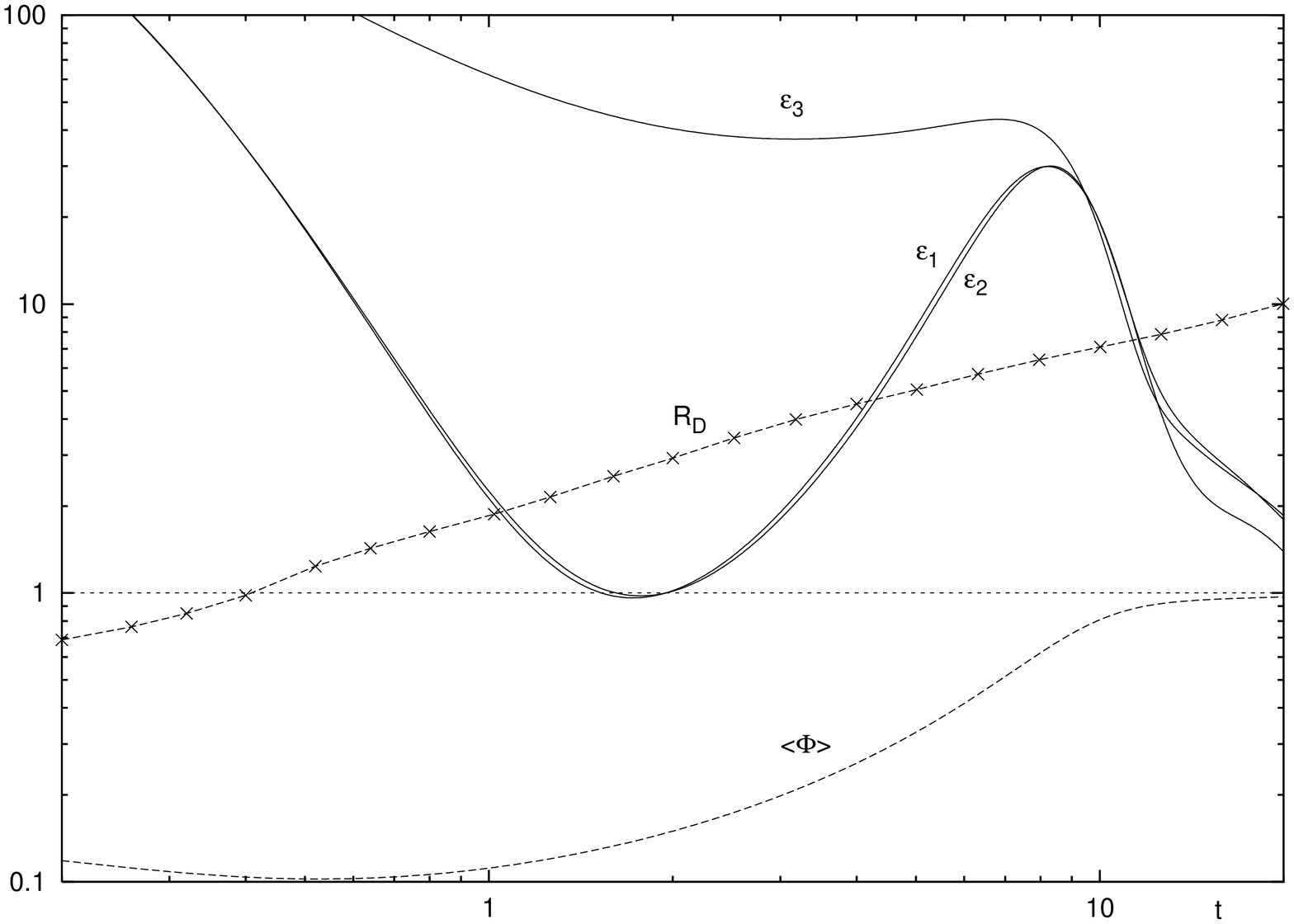}
\end{center}
\caption{ Angular correlation lengths $R_D$, average bag fields
$\langle\Phi\rangle$, and the rates $\epsilon_i$ of energy loss
(\ref{epsi}) carried by each of the three components $i=1,2,3$ of the
chiral field, as functions of time~$t$ after a sudden quench, on
a 30$\times$30 (left) and an 80$\times$80 (right) grid (plain linear
$\sigma$-model, no constraint on $B$).}
\label{eps}
\end{figure}

Let us at first look at the rates $\epsilon_i$ (\ref{epsi}) of energy
carried away 
by the field components $i=1,2,3$ for the pure linear $\sigma$-model with
no explicit symmetry breaking, $m_\pi=0$. The direction of spontaneous
symmetry breaking then is fixed by the surrounding vacuum alone, which is
imposed through the condition that the chiral field is kept at
$\Bphi=(0,0,1)$ at all times on the boundary of the lattice. 
For these events the roll-down phase where the average bag field $\langle
\Phi \rangle$ grows from a small value to saturation $\langle
\Phi \rangle \approx 1$ covers the time intervall from $t\approx 1$ to
$t\approx 10$ in units of the relaxation time $\tau$. During this time
interval the correlation lengths $R_D$ grow from values near
$R_D\approx 1$ to $R_D \approx 10$. These growth rates are quite
independent of the lattice size $L$, such that the average number of
disordered domains by the end of the roll-down phase is of the order of
$(L^2/(\pi R_D^2)$.

In figs.\ref{eps} we show for some arbitrarily selected individual event 
the rates $\epsilon_i$ of emitted energy for two lattice sizes, $L=30$ and
$L=80$. Due to the symmetry breaking effect of the surrounding vacuum,
emission into the $i=3$ $\sigma$-direction dominates the pionic $i=1,2$
emission rates during most of the roll-down phase by one to two orders
of magnitude. Naturally, this feature is more pronounced in the 
smaller lattice. Only after completion of the roll down, when the
absolute emission rates are getting very small, the average rates
become again comparable for $i=1,2,3$. While $\epsilon_3$ is essentially
monotonously decreasing during the whole ordering process, the pionic
emission occurs in two well-separated pulses. An early, short pulse
with high intensity is emitted for times $t < 1$ and it is
characterized by equipartition $\epsilon_1 = \epsilon_2$. This is as
expected because at those times the angular correlation lengths $R_D$ 
are near or less than 1. 
The second pulse appears during the second half of the
roll-down phase; its maximum is at least one order of magnitude smaller
than that of the first pulse and it is reached near $\langle \Phi
\rangle \approx 
0.8$. For the larger lattice also this second pulse is characterized
by equipartition for both isospin components. However, for the smaller
lattice, a significant difference between $\epsilon_1$ and $\epsilon_2$
can be observed. It develops with the onset of the roll down and
extends over the whole second pulse of pion emission and thus
constitutes a possible DCC signature. Unfortunately, it is just in the
small lattice where the pionic signal is strongly dominated by 
$\sigma$-emission. (Counting particles instead of energy, this is,
however, compensated by the mass ratio).

\begin{figure}[h]
\begin{center}
\includegraphics[width =6cm,height=8cm,angle=-90]{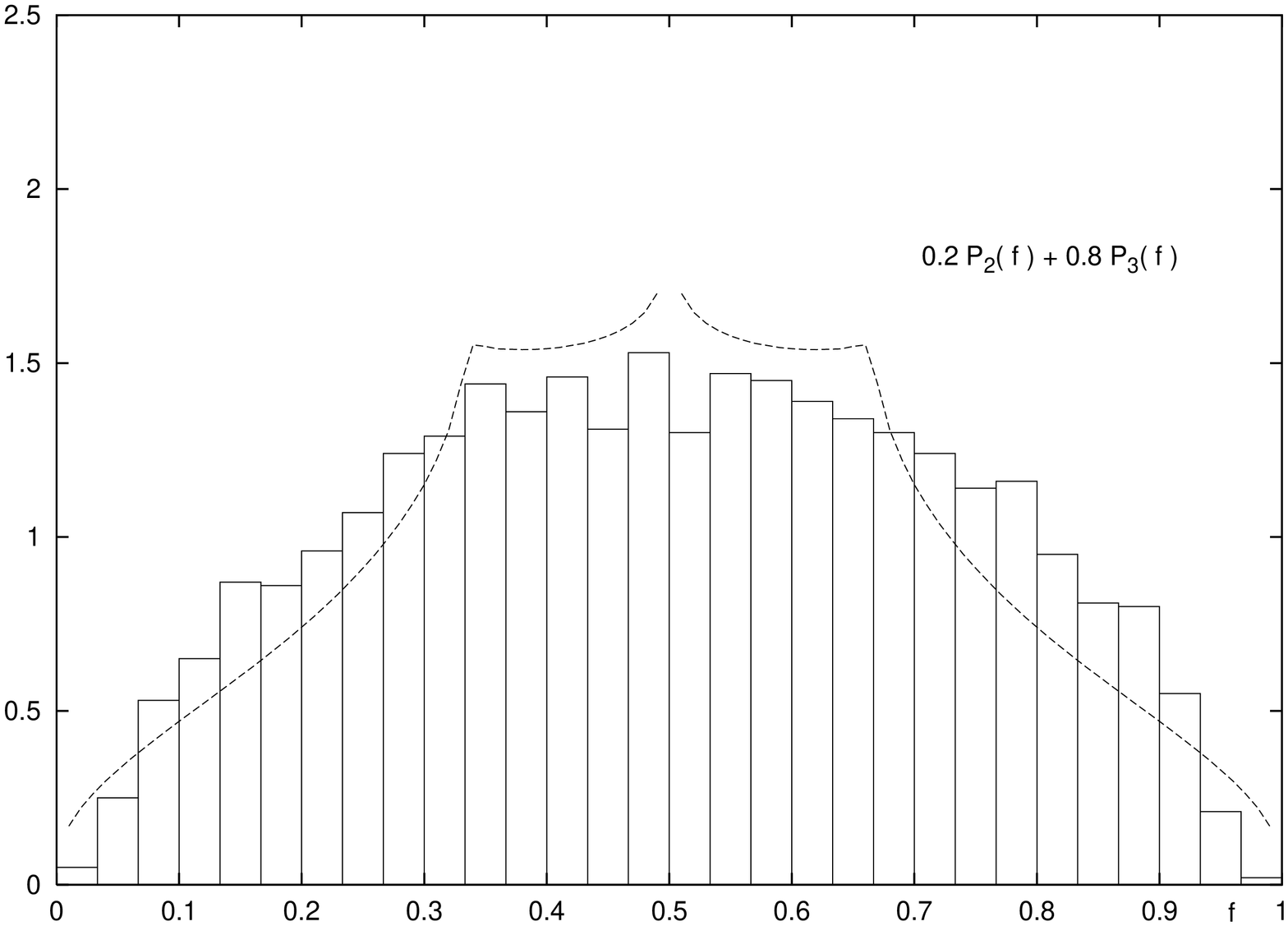}
\includegraphics[width =6cm,height=8cm,angle=-90]{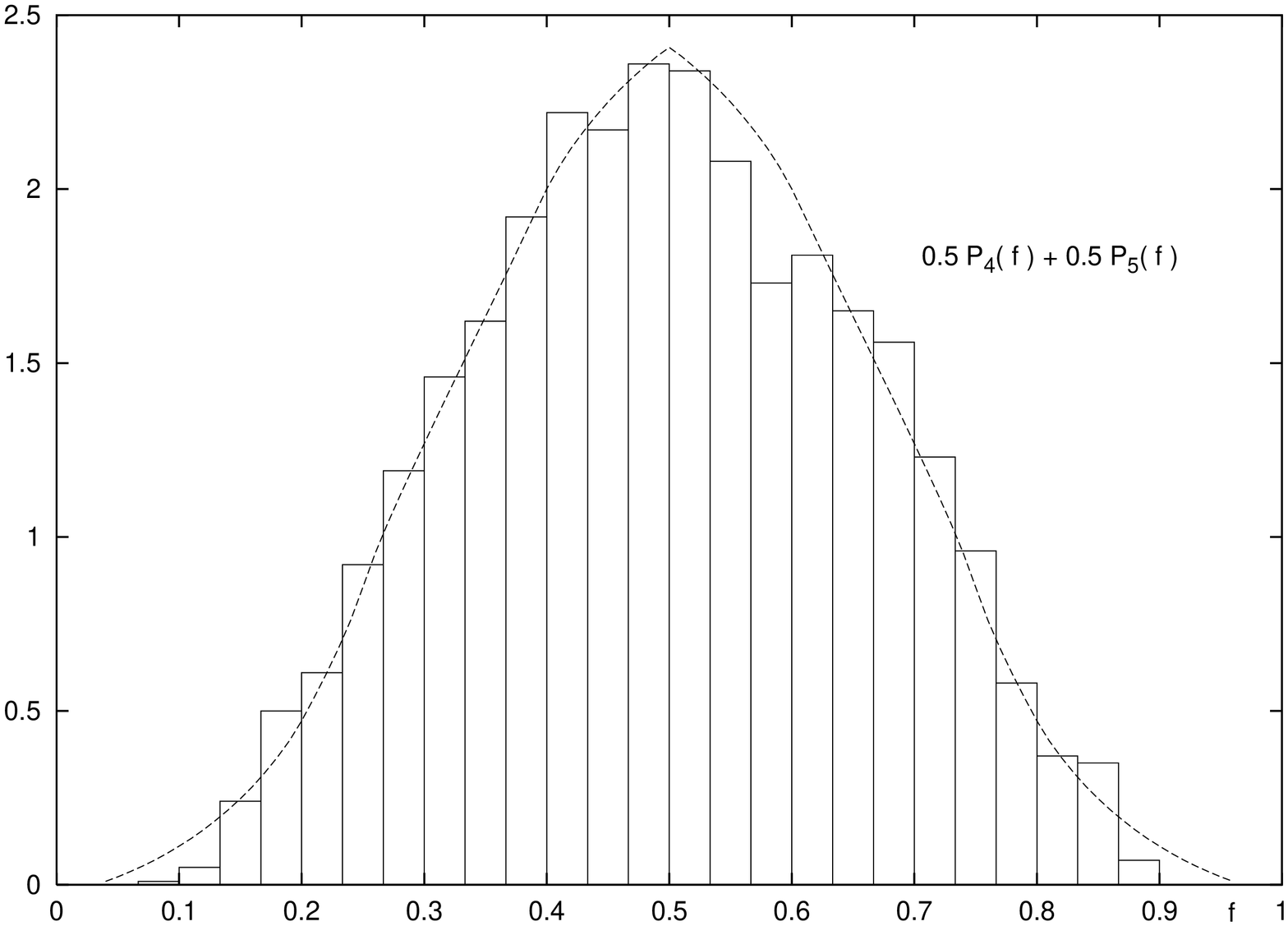}
\end{center}
\caption{Probability distributions $P(f)$ for the abundance ratios
$f_1(t)$ (\ref{fi}) of DCC pions extracted from an ensemble of 3000
events at roll-down time $t_{RD}$. For comparison, suitable mixtures 
of $P_\nu(f)$ selected from the approximation (\ref{Pnu}) are shown 
by the dashed lines.
The events evolve after a sudden quench on a 30$\times$30 grid, with no
explicit symmetry breaking, according to the plain linear
$\sigma$-model without constraint (left), 
and for the full model with $B=0$ (right).}
\label{Pf30}
\begin{center}
\includegraphics[width =6cm,height=8cm,angle=-90]{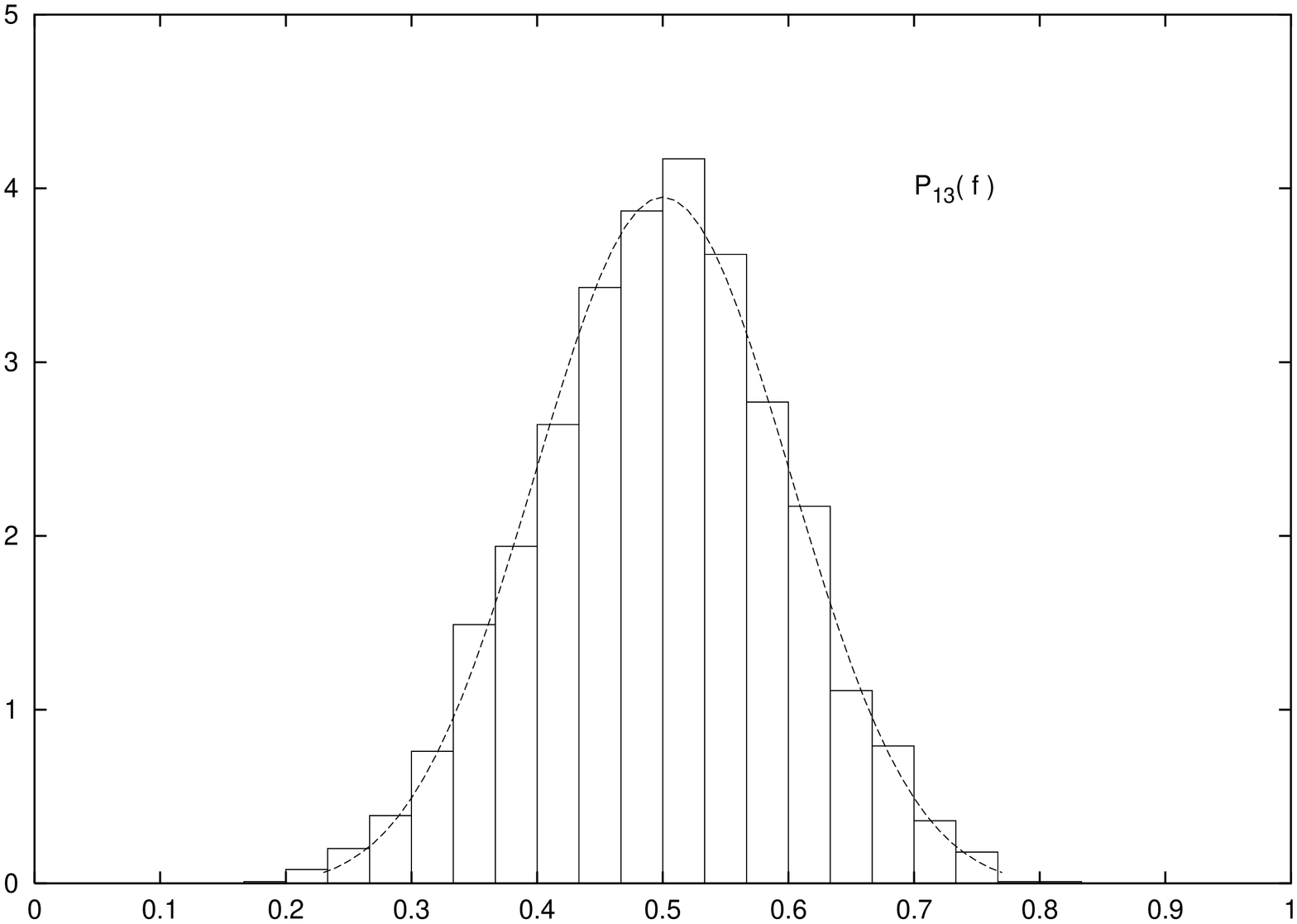}
\includegraphics[width =6cm,height=8cm,angle=-90]{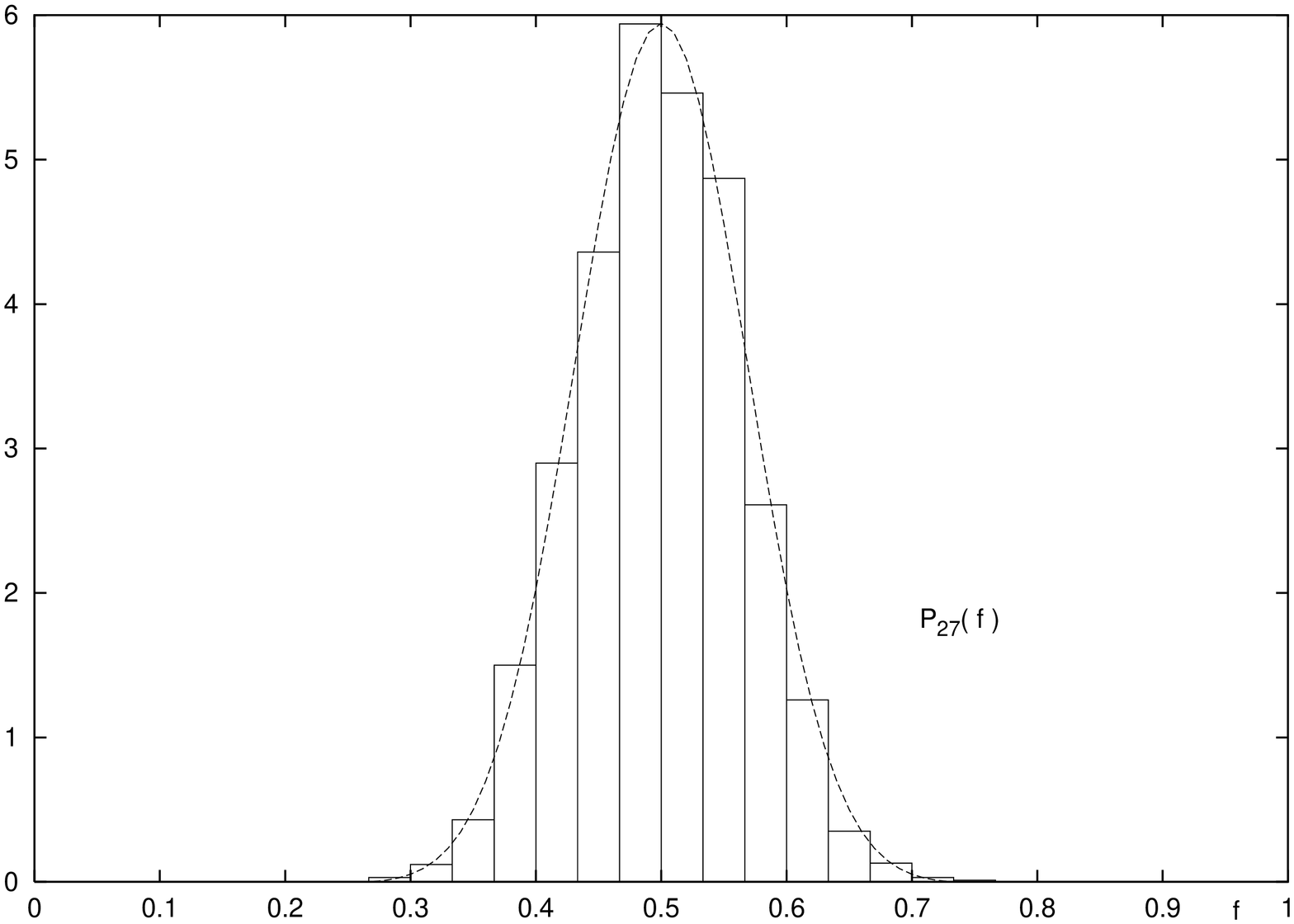}
\end{center}
\caption{ The same as in fig.\ref{Pf30}, but for a 60$\times$60 grid. }
\label{Pf60}
\end{figure}

It is instructive to analyse the two pion pulses in terms of the
TDGL-equation (\ref{eom}) which relates (in the pure linear
$\sigma$-model with $m_\pi=0$)
$\dot{\Phi_i}$ to $\Delta\Phi_i
-(\lambda/\ell^2) \left( \Phi^2-f^2 \right) \Phi_i$. As expected, the
first pulse is mainly due to the angular ordering effectuated by
$\Delta\Phi_i$, while the second pulse is dominated by the potential
term. So, the simple approximation $\epsilon_i \sim \langle \left(
\Phi^2-f^2 \right)^2 \Phi_i^2 \rangle $ provides the essential features
of the second maximum, (the 'DCC pions'). We may even go further and
replace $\Phi^2$ by its lattice average $\langle \Phi^2 \rangle$, such
that the abundance ratios (\ref{fi}) for DCC pions are approximated by 
\be
\label{fpi}
f(\pi_i) \approx \frac{\langle \Phi_i^2 \rangle}{\langle \Phi_1^2
\rangle + \langle \Phi_2^2 \rangle}.
\ee  
In this approximation we recover the commonly used prescription~\cite{Anselm}
to take the strength of pion emission proportional to the square of the
classical field components. (Evidently, this works only for ratios, and
only for pions in the second peak). Based on this approximation, we may
derive simple expressions for the probabilities $P_\nu(f)$, to observe
in a large ensemble of configurations with $\nu$ disoriented domains
the ratios $f(\pi_i)$  of DCC pion multiplicities (see Appendix).

\begin{figure}[h]
\begin{center}
\includegraphics[width =7cm,height=8cm,angle=-90]{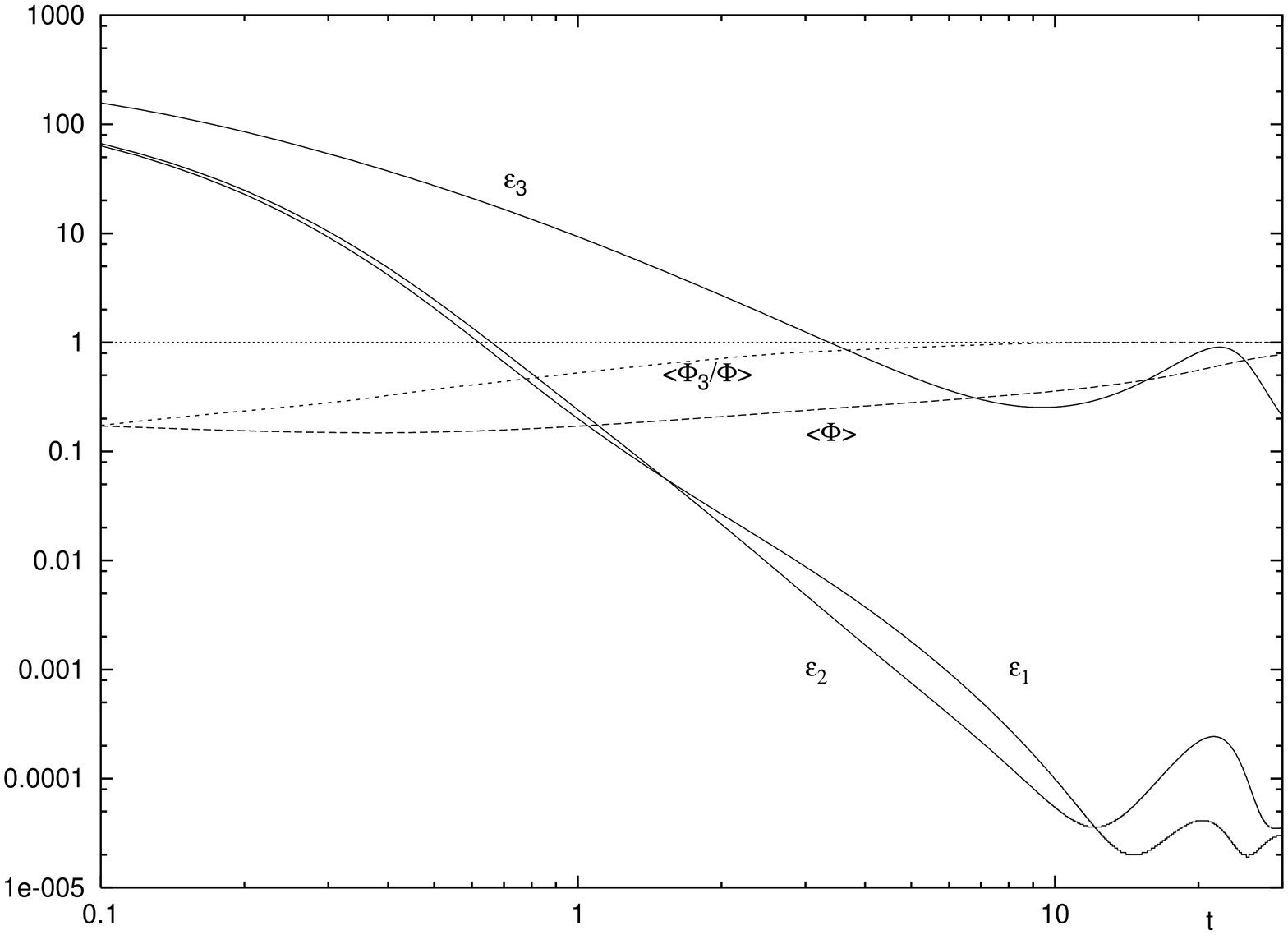}
\includegraphics[width =7cm,height=8cm,angle=-90]{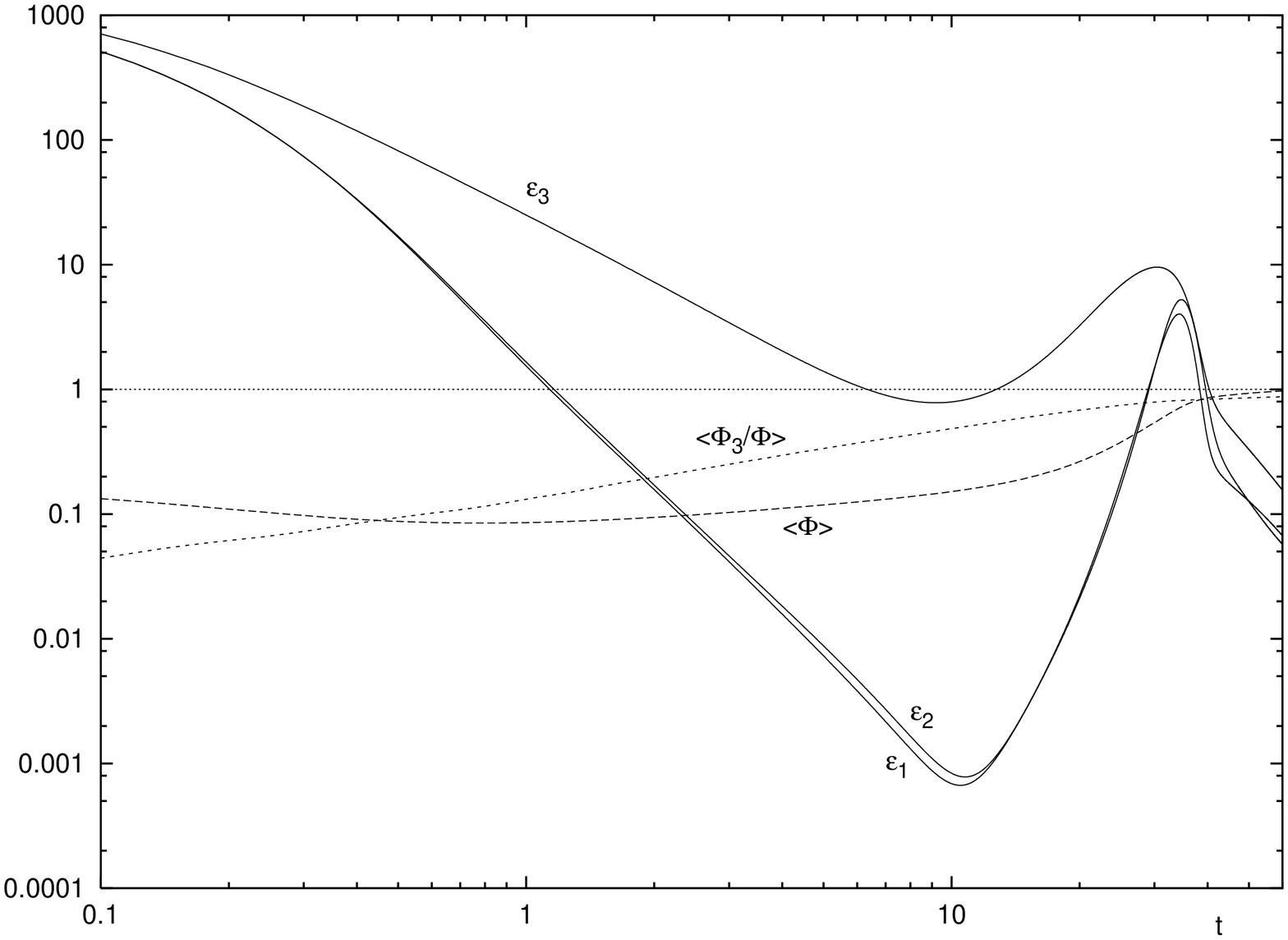}
\end{center}
\caption{Average bag fields $\langle\Phi\rangle$, 
averaged 3-components of the field unit vector $\langle\hphi_3\rangle$,
and the rates $\epsilon_i$ of energy loss 
(\ref{epsi}) carried by each of the three components $i=1,2,3$ of the
chiral field, as functions of time~$t$ after a very slow quench
($\tau_c=100$), on a 30$\times$30 (left) and an 80$\times$80 (right)
grid  (plain linear $\sigma$-model, no constraint on $B$).} 
\label{slow}
\end{figure}

As discussed in the previous chapter, addition of explicit symmetry
breaking through a small pion mass like $m_\pi=0.1$ speeds up the roll-down
phase and the aligning of $\Bphi$ into the 3-direction and thus shifts
the DCC pions peak to slightly smaller times without affecting the
ratios. However, the important difference is that for $m_\pi \not= 0$
the emission rate for
the 3-component is strongly enhanced relative to the DCC pions, such
that also for a large lattice ($L=80$) during and beyond the whole roll-down
phase $\epsilon_3$ dominates the DCC pions by more than an order of
magnitude (for $m_\pi=0.1$). (For the small lattice ($L=30$)
$\epsilon_3$ dominates by two orders of magnitude).

The fourth-order Skyrme term plays a minor role during these early
stages of the ordering process. Its size stabilizing effect for the emerging
solitons affects the late stages after completion of the roll down.
Still, as we have discussed at the end of sect.4, already the emergence
of stable extended structures during the roll-down phase
presents an obstacle against the growth of large aligned domains 
and thus it counteracts large differences in the ratios $f_1$ and
$f_2$ of DCC pions. This becomes apparent by comparing the probability
distributions $P(f)$ obtained from a large ensemble of 3000 events on a 
30$\times$30 lattice (fig.\ref{Pf30}), for the plain unconstrained
linear $\sigma$-model, and the full model (\ref{lag}) with constraint
$B=0$. We sample into 30 bins the
ratios $f(\pi_1)$ extracted from 3000 evolutions at the time when 
$\langle \Phi \rangle$ grows beyond 0.8, which approximately 
marks the peak in the DCC-pions pulse. Comparing with linearly
mixed $P_\nu(f)$ taken from (\ref{Pnu}) for two neighbouring values of
$\nu$, we may conclude 
that the presence of stable textures increases the average number of
disoriented domains present at that point in time
(for the fixed lattice size $L=30$) from 2-3 to 4-5. As we have
discussed above, this average number of $\nu$ increases with $L^2$.
Fig.\ref{Pf60} shows the corresponding distributions for 3000 events
on a $L=60$ lattice. The distribution obtained from the linear
$\sigma$-model is very nicely reproduced by $P_\nu(f)$ with $\nu=13$,
while the full and constrained to $B=0$ evolutions are compatible with
$\nu= 27$, which again shows the strong obstruction of domain growth 
through the creation of stable textures. 
Still, it should be remembered that the 
$P_\nu(f)$ of (\ref{Pnu}) which we use for comparison are derived under
the rather poor approximation (\ref{fpi}) and the assumption that all
domains are of equal size.  

To illustrate the influence of the cooling rate for slow quenches
figs.\ref{slow} show two evolutions with cooling time $\tau_c=100$ as
discussed in sect.5. With this choice the cooling time is 
large as compared to the roll-down time after a sudden quench.
The decisive feature of the evolution on a small 30$\times$30 lattice
is that the saturation of $\langle \Phi_3 \rangle$ is almost completed 
before the begin of the roll-down of $\langle \Phi \rangle$.
Consequently, the DCC pion pulse which still occurs during the late
roll-down phase, is suppressed by almost four orders of magnitude as
compared to $\sigma$-emission $\epsilon_3$. On this small lattice 
the disoriented domains show up in the pion emission ratios, while on
a larger (80$\times$80) grid, the saturation of $\langle \Phi_3 \rangle$
is delayed, the height of the DCC pion pulse increased, but it is
characterized by equipartition between $\epsilon_1$ and $\epsilon_2$.
So, also in this respect the sudden quench provides the better
conditions for a DCC signal.

\section{Conclusion}

In this paper we have investigated the phase ordering 
of a 2D-$O(3)$ model under dissipative dynamics. The peculiarity of the
underlying lagrangian is that it stabilizes extended textures through a
fourth-order current-current coupling. This is introduced to get an
idea how emerging topological solitons with stable sizes interact with
the growth of aligned domains. The winding number can be kept fixed to
simulate in this model the analogy of baryon number conservation which 
is an important feature of models of hadronic systems. Two parameters
characterize 
the model: One fixes the spatial size scale of the stabilized 'baryons',
the other determines the depth of the bag profile, into which the
winding (baryon-) density is confined. Shallow bags correspond 
basically to the nonlinear version of the model, deep bags
realize the unbroken chiral symmetry in the interior of baryons, 
as it would be expected in hadronic systems.

Apart from studying the influence of different ways of symmetry
breaking, and the effects of different quench rates on the ordering
observables, we are especially interested in the possibility to
find DCC signatures in the emitted particle radiation. All results are
based on the assumption that the relaxation proceeds on a sufficiently
slow time scale, such that it provides an adiabatically changing source
for the emitted radiation. Therefore its motion is treated 
classically, and the amount of radiated energy is simply obtained from
the classical energy loss of the source. Quantum field-theoretic
amplitudes enter only into the relaxation constant $\tau$ which, however,
has not been determined, but simply used as time unit for the ordering
dynamics. 

From the results obtained in this framework we may draw the following
conclusions: The classical Allen-Cahn growth law $R_D\sim t^{0.5}$ for
disoriented domains is not seriously disturbed by the stabilizing terms
in the lagrangian and the emerging stable textures, although there are
characteristic deviations from it during the roll-down phase where the
separation into domains of false vacua and deep bags mainly takes
place. The factor in front of the $t^{0.5}$ power law, however, is
decreased as compared to the pure unconstrained linear $\sigma$-model,
such that the average size of disoriented domains after completion of
the roll-down is reduced. The roll-down time varies with the two
parameters of the model essentially as expected from simple scaling
arguments. It represents the crucial characteristic of the ordering
process, it is not very sensitive to small explicit symmetry breaking
and varies with the quench rate in the expected way. Explicit symmetry
breaking strongly affects the saturation time of the average
orientation, the factor in front of the exponential approach to 
saturation depends severely on the (random) initial conditions of each
individual event. The number of baryons (plus antibaryons) finally surviving
the ordering process is determined by the time of the onset of the roll
down. The angular correlation length at that time measures the average
distance of regions with high winding density which develop into separate
stable structures with integer winding number during and after the roll
down phase. Therefore slow quenches lead to low baryon density.

The intensities of emitted radiation in our approach is determined by
the square of the time-derivative of the classical field, averaged
over the spatial volume (or area), and not by the square of the classical
field itself. This allows to identify two well-separated radiation
pulses for the transverse ('pionic') components of the chiral field.
The later one, emitted towards the end of the roll-down phase 
carries the signature of the disoriented domains established by that
time, so it constitutes the DCC signal. 
With the growth law of the angular correlations fixed, the average size
of disoriented domains at the time of completion of the roll down is
also fixed. 
Therefore the average number of disoriented domains on an $L \times L$ 
lattice which contribute to the DCC signal increases like $L^2$,
and deviations from equipartition in the intensities carried by
the different isospin directions quickly vanish with increasing $L$.
On the other hand, for small $L$, the pion signal is buried under a
large pulse of emission in longitudinal ($\sigma$-)direction, because
the explicit symmetry breaking effect of the surrounding true vacuum
is strong for small lattice sizes. 

The broadening of the probability distributions for the pionic 
intensity ratios constitutes the DCC signature. We have extracted it
from large ensembles of events under the assumption that these pions,
emitted at a specific point in time (or averaged over a short time
interval near the roll-down time) can be observed separately from
the bulk of other pions and $\sigma$'s emitted. We find that in those 
cases where a noticeable broadening occurs, the relative amount of
energy carried away in the DCC pulse, is suppressed by orders of
magnitude. Of course, quantitative statements require knowledge of
the parameters in the model, cooling rates, physical size of the
lattice as compared to correlation length, baryon size.
In the present model, all these conditions remain more or less
arbitrary, and we have chosen them at will to illustrate the main
features. However, for application to the ordering dynamics of a 
hadronic chiral $O(4)$-field, we have quite specific ideas about
the relevant parameters in the effective actions. 
The exponent in the classical Allen-Cahn scaling law is the same
for (2+1)D-$O(3)$ and (3+1)D-$O(4)$ models; the homotopy groups
$\pi_2(S^2)$ and $\pi_3(S^3)$ are isomorph, both characterized by
integer winding numbers. So, apart from differences in 
dimension-dependent cooling rates and temperature-dependence of
the condensate $f(T)$, we do not expect dramatic qualitative
differences in the ordering features, and we hope that
extending the 
present considerations to the (3+1)D-$O(4)$ model will serve to sharpen
our expectations with respect to the observability of DCC signals
in hadronic processes.

\acknowledgements
The author appreciates helpful discussions with H.Walliser.

\begin{appendix}
\section{Probability distributions}
\label{A}
\begin{figure}[h]
\centering
\includegraphics[width=9cm,height=12cm,angle=-90]{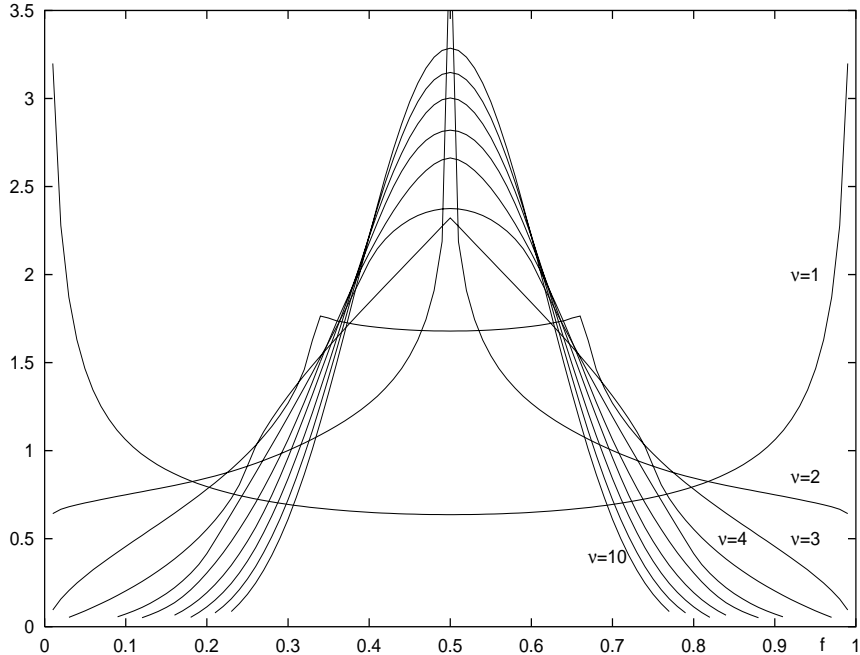}
\caption{The probability distributions $P_\nu(f)$ for
observing the fraction $f$ of pions with a given isospin
component $i=1$ or $i=2$ in an ensemble of configurations with $\nu$
disoriented domains of size $1/\nu$, according to (\ref{Pnu}). }
\label{Pfig}
\end{figure}

Consider a configuration where the classical field is uniformly
aligned into some direction, thus forming one single disoriented
domain, with the field 
components ($i=1,2,3$) parametrized as 
\be
\label{orient}
\Phi_{i=1,2,3}=\Phi \large(\sin\theta \cos\phi,\sin\theta
\sin\phi,\cos\theta \large) .
\ee
Then, within the approximation (\ref{fpi}),
we find for the fraction of pions emitted with isospin component
$i=1$, relative to the number of all pions  $(i=1,2)$ 
\be
\label{fp1}
f(\pi_1)  = \cos^2\phi. 
\ee
In an ensemble of events
where all orientations of $\Phi_{1,2}$ are equally probable
the ensemble average $\langle f \rangle$ of $f(\pi_1)$
is, of course, $\langle f \rangle=1/2$.
The probability $P(f)$ to find in one event of that ensemble the
fraction $f$ (for pions of either kind $i$)
then is obtained from
\be
\langle f \rangle=\frac{1}{2\pi}\int_0^{2\pi}f d\phi =\int_0^1 f P(f) df
\ee
as
\be
\label{Pf}
P(f)=\frac{1}{\pi \sqrt{f(1-f)}}.
\ee
(Note the difference to the standard chiral $O(4)$-model, where
$P(f)=1/(2 \sqrt{f})$.)
For $\nu$ such domains of size $1/\nu$ the probability $P_\nu(f)$ to
find the fraction $f$ in a given event then is
\be
\label{Pnu}
P_\nu(f)=\int_0^1...\int_0^1 \;\delta\left( f-\frac{1}{\nu}
(f_1+...+f_\nu)\right) P(f_1)...P(f_\nu)\; df_1...df_\nu
\ee
with $P(f)$ given by (\ref{Pf}). Some of these functions, for values
from $\nu=1$ to $\nu=10$, are shown in fig.\ref{Pfig}. For increasing
$\nu$ they are, of course, more and more sharply peaked around $f=1/2$.

\end{appendix}


\begin{thebibliography}{99}
\itemsep=0.5cm

\bibitem{Bray} For a review, see A.J. Bray, {\it{Adv.Phys.}} {\bf 43}
(1994) 357.

\bibitem{Raja} K. Rajagopal and F. Wilczek, {\it{Nucl.Phys.}} {\bf B404}
(1993) 577; S. Gavin, A. Gocksch, and R.D. Pisarski, 
{\it{Phys.Rev.Lett.}} {\bf 72} (1994) 2143; G. Amelino-Camelia, 
J.D. Bjorken, and S.E. Larsson, 
{\it{Phys.Rev.}} {\bf D56} (1997) 6942; 
M. Asakawa, Z. Huang, and X.-N. Wang, {\it{Phys.Rev.Lett.}} {\bf 74} 
(1995) 3126; 
A.K. Chaudhuri\-, hep-ph/0007332;
 J. Serreau, {\it{Phys.Rev.}} {\bf D63} (2001) 054003. 


\bibitem{Randrup} D. Boyanovsky, D.-S. Lee, and A. Singh, 
{\it{Phys.Rev.}} {\bf D48} (1993) 800;
D. Boyanovsky, H.J. de Vega, and R. Holman, {\it{Phys.Rev.}}
{\bf D51} (1995) 734;
F. Cooper, Y. Kluger, E. Mottola, and J.P. Paz,  {\it{Phys.Rev.}}
{\bf D51} (1995) 2377;
M.A. Lampert, J.F. Dawson, and F. Cooper, {\it{Phys.Rev.}}
{\bf D54} (1996) 2213;
J. Randrup, {\it {Phys.Rev.Lett.}} {\bf 77} (1996) 1226; 
{\it{Phys.Rev.}} {\bf D55} (1997) 1188;
{\it{Heavy Ion Phys.}} {\bf 9} (1999) 289; {\it{Phys.Rev.}} {\bf C62}
(2000) 064905. 


\bibitem{Anselm} A.A. Anselm, {\it{Phys.Lett.}} {\bf B217} (1989) 169; 
A.A. Anselm and M.G. Ryskin, {\it{Phys.Lett.}} {\bf B266}  (1991) 482;
J.P. Blaizot and A. Krzywicki, {\it{Phys.Rev.}} {\bf D46}  (1992) 246;
{\it{Phys.Rev.}} {\bf D50} (1994) 442; {\it{Acta Phys.Polon.}} {\bf B27}
(1996) 1687; J.D. Bjorken, {\it{Int.J.Mod.Phys.}} {\bf A7} 
(1992) 4189; {\it{Acta Physica Polonica}} {\bf B23} (1992) 561,
 {\bf B28} (1997) 2773; K.L. Kowalski\- and C.C. Taylor, SLAC-PUB-92-6; 
K.L. Kowalski, J.D. Bjorken, and C.C. Taylor, SLAC-PUB-6109 (1993); 
S. Gavin, {\it{Nucl.Phys.}} {\bf A590}  (1995) 163c.  




\bibitem{Skyrme} T.H.R. Skyrme, {\it{Proc.R.Soc.}} {\bf A260} 
 (1961) 127;\\  
E. Witten, {\it{Nucl.Phys.}} {\bf B223}  (1983) 422,433.  


\bibitem{Kibble} T.W.B. Kibble, {\it{J.Phys.}} {\bf A9}  (1976) 1387;\\  
N.H. Christ, R. Friedberg, and T.D. Lee, 
{\it{Nucl.Phys.}} {\bf B202}  (1982) 89.

\bibitem{Ellis} T.A. DeGrand, {\it{Phys.Rev.}} {\bf D30}  (1984) 2001;\\  
J. Ellis and H. Kowalski, {\it{Phys.Lett.}} {\bf B214} 
 (1988) 161; J. Ellis, U. Heinz, and H. Kowalski, {\it{Phys.Lett.}} 
{\bf B233}  (1989) 223; J. Ellis, M. Karliner, and H. Kowalski, 
{\it{Phys.Lett.}} {\bf B235}  (1990) 341; J. Dziarmaga and M. Sadzikowski,
{\it{Phys.Rev.Lett.}} {\bf 82}  (1999) 4192.

\bibitem{Asa} M.Asakawa, H. Minakata, and B. Mueller, 
{\it{Phys.Rev.}} {\bf D58}  (1998) 094011.

\bibitem{Biro} T.S. Biro and C. Greiner, {\it{Phys.Rev.Lett.}} {\bf 79}
(1997) 3138; Z. Feng, D. Molnar, and L.P. Csernai, hep-ph/9702246.


\bibitem{Zak}  A.D. Rutenberg, {\it{Phys.Rev.}} {\bf
E51} (1995) R2715;\\
M. Zapotocky and W.J. Zakrzewski,  {\it{Phys.Rev.}} {\bf
E51} (1995) R5189;
A.D. Rutenberg, W.J. Zakrzewski, and M. Zapotocky,
{\it{Europhys.Lett.}}  {\bf 39}  (1997) 49;\\
G.J. Stephens, {\it{Phys.Rev.}} {\bf D61}  (2000) 085002.    

\bibitem{Sondhi}
S.L. Sondhi, A. Karlhede, S.A. Kivelson, and E.H. Rezayi,
{\it{Phys.Rev.}}  {\bf B47}  (1993) 16419. 

\bibitem{Ho} G. Holzwarth, {\it{Nucl.Phys.}} {\bf A672} (2000) 167.


\bibitem{HoKl} G. Holzwarth and J. Klomfass, {\it{Phys.Rev.}} {\bf D63}
 (2001) 025021. 

\bibitem{AlCa} S.M. Allen and J.W. Cahn, {\it{Acta metall.}}
{\bf 27} (1979) 1085.

\bibitem{BrayII} J.A.N. Filipe, A.J. Bray, and S. Puri, 
{\it{Phys.Rev.}} {\bf E52} (1995) 6082.

\end{thebibliography}
\end{document}